\documentclass[fleqn,usenatbib,usedcolumn]{mnras}

\usepackage{newtxtext,newtxmath}

\usepackage[T1]{fontenc}
\usepackage{ae,aecompl}

\usepackage{graphicx}	
\usepackage{amsmath}	
\usepackage{amssymb}	
\usepackage[usenames,dvipsnames]{xcolor}

\usepackage{soul}

\title[Bright white dwarfs for high-speed photometry]{Multi-band photometry and spectroscopy of an all-sky sample of bright white dwarfs}

\author[R. Raddi et al.]{R. Raddi$^{1}$\thanks{E-mail: r.raddi@warwick.ac.uk}, N. P. Gentile Fusillo$^{1}$, A. F. Pala$^{1}$, J. J. Hermes$^{2}$\thanks{Hubble Fellow}, B. T. G\"{a}nsicke$^{1}$
\newauthor P. Chote$^{1}$, M. A. Hollands$^{1}$, A. Henden$^{3}$,  S. Catal\'an$^{4}$, S. Geier$^{5}$, D. Koester$^{6}$, 
\newauthor U. Munari$^{7}$, R. Napiwotzki$^{8}$, P.-E. Tremblay$^1$ \\
$^{1}$Department of Physics, University of Warwick, Gibbet Hill Road, Coventry CV4 7AL\\
$^{2}$Department of Physics and Astronomy, University of North Carolina, Chapel Hill, NC-27599, USA\\
$^{3}$AAVSO, 49 Bay State Rd Cambridge, MA 02138, USA\\
$^{4}$School of Physics, National University of Ireland, Galway, Ireland\\
$^{5}$Institute for Astronomy and Astrophysics, Eberhard Karls University, Sand 1, D 72076 T\"ubingen, Germany\\
$^{6}$Institut f\"ur Theoretische Physik und Astrophysik, Universit\"at Kiel, 24098, Kiel, Germany\\
$^{7}$INAF-Astronomical Observatory of Padova, I-36012 Asiago (VI), Italy\\
$^{8}$Centre for Astrophysics Research, STRI, University of Hertfordshire, College Lane, Hatfield AL10 9AB, UK}

\date{Accepted 2017 August 28. Received 2017 August 28; in original form 2017 February 27}

\pubyear{2017}

\begin{document}
\label{firstpage}
\pagerange{\pageref{firstpage}--\pageref{lastpage}}
\maketitle

\begin{abstract}
The upcoming NASA {\em Transiting Exoplanet Survey Satellite} ({\em TESS}) will  obtain space-based 
uninterrupted light curves for a large sample of bright white dwarfs distributed across the entire 
sky, providing a very rich resource for asteroseismological studies and the search for transits from 
planetary debris. We have compiled an all-sky catalogue of ultraviolet,
optical, and infrared photometry as well 
as proper motions, which we propose as an essential tool for the preliminary identification and characterisation of potential targets.  
We present data for 1864 known white dwarfs and 305 
high-probability white dwarf candidates brighter than 17\,mag. We describe the spectroscopic 
follow-up of 135 stars, of which 82 are white dwarfs and 25 are hot subdwarfs. The new confirmed
stars include six pulsating white dwarf candidates (ZZ Cetis), and nine white 
dwarf binaries with a cool main-sequence companion. We identify one star with 
a spectroscopic distance of only 25\,pc from the Sun. 
Around the time {\em TESS} is launched, we foresee that all white dwarfs in this sample
will have trigonometric parallaxes measured by the ESA {\em Gaia} mission next year.\end{abstract}




\section{Introduction}
White dwarfs are the final evolutionary stage for the vast majority of single stars.  
Characterised by long, steady cooling rates, white dwarfs can be used to infer properties of  
the Milky Way like the the star formation history \citep[][]{Tremblay14}, the age of the disc  
\citep[][]{Winget87, Oswalt96}, or their contribution to baryonic dark matter 
\citep[][]{Reid01, Pauli03}. 

Asteroseismology is a powerful tool for probing the internal structure of white dwarfs,
through the  study of multi-periodic variability that occurs during short-lived instability phases on the white dwarf cooling sequence  \citep[][]{Fontaine08, Winget08, Althaus10}. 
Over the last few years, the NASA {\em Kepler} mission \citep[][]{Borucki10} and its re-purposed two-wheeled {\em K2} mission \citep[][]{Howell14}
have enabled high-precision asteroseismology of white dwarfs \cite[][]{Hermes11, Ostensen11, Greiss16}. Its successor, the NASA {\em
Transiting Exoplanet Survey Satellite} \citep[{\em TESS};][]{Ricker15}, will acquire all-sky times-series photometry for 
at least two years,
searching for exoplanet transits and  enabling asteroseismology for many classes of stars \citep[][]{Campante16}.  Roughly 200\,000 selected
targets, including bright white dwarfs, may be observed with  2-min cadence. Full-frame images 
will be taken every 30 min. The mission will survey areas of sky with one-month to one-year coverage, 
depending on their ecliptic latitudes. 

{\em TESS} is poised to return improved statistics building upon the original {\em Kepler} 
mission and its ongoing {\em K2} reincarnation. In particular, the large sky coverage of {\em TESS}
will be excellent for the discovery of rare phenomena of which {\em Kepler} has so far
provided us with few, or even unique examples, e.g. the first direct detection 
of a disintegrating asteroid at a white dwarf \citep[][]{Vanderburg15} or the
identification of stochastic outbursts in cool pulsating white dwarfs \citep[]{Bell15, Hermes15, Bell16}.  
{\em TESS} is expected to be launched in 2018 March--July, 
which is just before or at the same time of the second data release of the ESA
{\it Gaia} mission \cite[DR2;][]{Gaia16a, Lindergren16}. 
Therefore, the need of an all-sky target list of bright white dwarfs is extremely urgent, 
especially for the southern ecliptic hemisphere, which will not have the benefit of {\em Gaia} selection. 
Bright white dwarfs can also be monitored by upcoming photometric missions, 
such as Evryscope \citep{Law15} or the Next Generation Transit Survey \citep[NGTS;][]{West16}.

Traditionally, the identification of  white dwarfs relies on their blue colours 
and relatively large proper motions. The availability
of accurate digital photometry and astrometry has greatly improved the efficiency of white dwarf selection 
\citep[e.g.][]{Kawka06, Kilic06, Limoges13, GentileFusillo15, Raddi16, GentileFusillo17}.  
Numerous efforts have been undertaken to improve the identification and
characterisation of bright and nearby white dwarfs \citep[][]{Sion14, Holberg16}, not only through spectroscopy and photometry
\citep[][]{Giammichele12}, but also including trigonometric parallaxes  \cite[e.g. DENSE
project\footnote{\url{http://www.DenseProject.com}};][]{Subasavage09}. 
So far, the white dwarf census is strongly biased towards the
northern hemisphere, which \citet{Limoges15} estimate to be 78 per cent complete at 40\,pc.  

In the following sections, we describe a collection of proper motions 
and  multi-band all-sky photometry for potential {\em TESS} white dwarf targets. 
Out of 2990 known white dwarfs brighter than 17~mag, we 
re-identify 1864 stars, i.e. 62 
per cent of the total. We discuss the properties of the sample and its completeness
via comparison with catalogues of spectroscopically 
confirmed white dwarfs and the Lowell Observatory proper motion survey \citep[][GD sample]{Giclas80}. 
We present photometry and proper motions for 305 new high-probability white dwarf candidates, 
and 33 white dwarf candidates in the GD sample. 
In the last part of the paper, we discuss the properties of 82 white dwarfs and 25 hot subdwarfs 
for which we acquired follow-up spectroscopy.

\section{The data}
\label{chap2}
\begin{figure*}
\centering
\includegraphics[width=\linewidth]{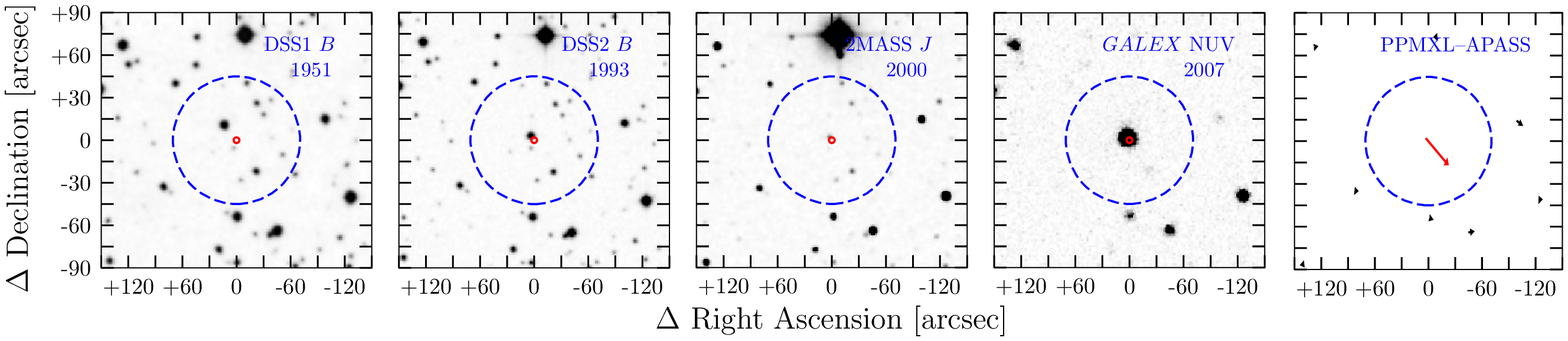}
\caption{In the first four panels, we show the evolving position of the newly confirmed white dwarf J0515+5021 (GD\,289). From left to right we display the cut-outs from the POSS-I and POSS-II blue plates, and the 2MASS\,$J$ and {\it GALEX} NUV images. The rightmost panel displays the PPMXL proper motion vector (in arcsec/yr, multiplied by 100  to improve the visualisation). In each panel, we plot the large cone-search of 45\,arcsec used to identify objects with large proper motions (in blue) and a 2-arcsec circle for the cross-match at the APASS position (in red).}
\label{f:cross_match}
\end{figure*}
The photometric identification of white dwarfs hotter than $\simeq 6000$\,K is very effective with colours sampling the flux across the Balmer jump, e.g. $ugr$ photometry of the Sloan Digital Sky Survey \citep[SDSS;][]{Girven11}.  Currently, there is no single all-sky survey covering this spectral region for magnitudes brighter than 17$^{\rm th}$ magnitude. A useful alternative is offered by the combination of  far- and near-ultraviolet photometry (FUV and NUV bands, centred at 1540 and 2270\,\AA, respectively) from the  {\it Galaxy Evolution Explorer}  \citep[{\it GALEX};][]{Morrissey07} with optical photometry from the AAVSO  Photometric All-Sky Survey \citep[APASS;][]{Henden14} in the Johnson $BV$ and Sloan $g'r'i'$ passbands. 

\citet{Bianchi11} discussed the properties of hybrid colour-colour diagrams of {\em GALEX} and SDSS photometry showing their potential at separating classes of Galactic and extragalactic sources. \citet{Limoges13} have used similar colour-combinations to select white dwarfs candidates in the northern hemisphere  from {\it GALEX}, SDSS, {\it Hipparcos} \citep[][]{vanLeeuwen07},  and {\it Tycho-2} \citep[][]{Hog00}, using proper motions from SUPERBLINK \citep[][]{Lepine05}. In the work described here, we used the PPMXL catalogue which includes positions and proper motions for 900 million stars derived from USNO-B\,1.0  \citep[][]{Monet03} and the Two-Micron All-Sky Survey \citep[2MASS;][]{Skrutskie06} astrometry in the International  Coordinate Reference System (ICRS).
In addition to the ultraviolet, optical, and near-infrared photometry, we also added mid-infrared data from the  {\em Wide-field Infrared Survey Explorer} \citep[{\em WISE};][]{Wright10}.  In Table\,\ref{t:surveys}, we list the wavelength coverage of the used set of filters.
\begin{table}
\centering
\caption{Photometric bands used in this work. Refer to 
the Spanish Virtual Observatory web pages$^{1}$ for the definition of
effective wavelength and width, $\lambda_{\rm eff}$ and $W_{\rm eff}$, respectively.\label{t:surveys}}
  \begin{tabular}{@{}lcrr@{}}
  \hline
Survey &  Filters & $\lambda_{\rm eff}$ & $W_{\rm eff}$ \\ 
 &   & [nm] & [nm] \\ 
  \hline
{\it GALEX} &  FUV & 154 & 25 \\
      &  NUV & 227 & 73  \\
APASS & $B$  & 430 & 84 \\
      & $g'$  & 464 & 116  \\      
      & $V$  & 539 & 87 \\
      & $r'$  & 612 & 111  \\
      & $i'$  & 744 & 104 \\
2MASS & $J$ & 1235 & 162 \\
      & $H$ & 1662 & 251 \\
      & $K_{\rm s}$ & 2619 & 162  \\
{\em WISE}     & $W1$ & 3353 & 663 \\
      & $W2$ & 4602 & 1042  \\
\hline
\multicolumn{4}{l}{1) \scriptsize \url{http://svo2.cab.inta-csic.es/theory/vosa4/}}
\end{tabular}
\end{table}
\subsection{The reference catalogue: APASS}
APASS data release 9 \citep[DR9;][]{Henden16} contains $\approx 61$\,million objects in the  $10^{\rm th}$--$17^{\rm th}$ magnitude range.  The survey uses two pairs of 20-cm twin telescopes: two in the north at Dark Ridge Observatory in New Mexico, and two in the south at the Cerro Tololo Inter-American Observatory. This setup allows automated, essentially simultaneous observations in the $BV$ and $g'r'i'$ bands. The pixel scale at detectors is 2.57\,arcsec, and photometry is extracted using a 15\,arcsec aperture in the North and 20\,arcsec in the South, due to differences in the focusing of telescopes. APASS\,DR9 is suggested to be currently complete down to $V = 16$, and it is expected to be complete down to $V=17$ at the end of the survey. The bright limit of APASS is of little concern here, as there are only two known white dwarfs brighter than 10\,mag (i.e. Sirius\,B, and 40\,Eri\,B; see next section). The astrometric accuracy is 0.15\,arcsec  and the photometric accuracy is 0.02\,mag at 15\,mag \citep[][]{Munari14}. The release notes\footnote{\url{https://www.aavso.org/apass}} suggest a careful use of APASS photometry in crowded areas, where the photometry extraction algorithm suffers from blending. Furthermore, the most recent measures of $B$, $g'$ magnitudes in the north are known to carry large random errors, due to a technical problem with the detectors. Nevertheless, APASS photometry was shown to be sufficiently accurate for tying other deeper surveys onto an absolute flux scale \citep[e.g.][]{Barentsen14, Drew14, Shanks15}.

Duplicated entries are present in APASS\,DR9 due to merging issues at the detector corners and for high proper motion objects. Thus, we generated a reduced dataset of unique sources within a separation of $< 1.25$\,arcsec, which corresponds to half a pixel of APASS detectors.  When more than one photometric value is identified in this internal cross-match process, we adopted the brightest detection as suggested in the APASS\,DR9 release notes. The so defined unique list of APASS sources contained a little less than 59 million objects.

\subsection{Matching with PPMXL, {\em GALEX}, and {\em WISE}}

To cross-correlate APASS with {\it GALEX} and {\em WISE} detections, we needed  to account for stellar proper motions. Thus, we first queried the PPMXL database via Table Access Protocol (TAP), using the Starlink Tables Infrastructure Library Tool Set \citep[{\sc stilts};][]{Taylor06}. The query was run by uploading the entire unique APASS dataset to the online service\footnote{PPMXL is hosted by the German Astronomical Virtual Observatory data center at \url{http://dc.zah.uni-heidelberg.de/}}, split in to chunks of 100\,000 rows each. For each APASS source, we identified all the PPMXL objects within 45\,arcsec. Then, assuming epoch 2013.0 for the APASS\,DR9 coordinates, we forward-propagated the coordinates of PPMXL sources using their proper motions. Finally,  we
identified  good matches by selecting objects that fall within 2\,arcsec from the APASS sources.
The APASS-PPMXL cross-match produced $\approx 45$\,million objects.

The next step was to cross-match the APASS-PPMXL sources with the {\it GALEX}\,DR6+DR7 database\footnote{{\it GALEX}\,DR6+DR7 is hosted by the Mikulski Archive for Space Telescopes (MAST) at  \url{http://galex.stsci.edu/GR6/}}, accessing the Catalog Archive Server Jobs System (CasJobs) via Simple Object Access Protocol (SOAP). In CasJobs we first queried the PhotoObjAll table to search for all the nearby sources within 45-arcsec from the position of our APASS-PPMXL sources; 
then, we identified the corresponding observing dates in the photoExtract table to propagate APASS coordinates to the corresponding {\it GALEX} epochs, again using the PPMXL proper motions. The cross-match with {\it GALEX} reduced further the number of APASS-PPMXL sources with proper motions and NUV photometry, leaving $\approx19$\, million 
detections (i.e. about 15~million unique sources).  

As last step, we cross-matched the APASS-PPMXL source list with {\em WISE} via TAP. Again, we shifted the APASS coordinates backwards in time by taking into account the epoch difference between APASS and {\em WISE}. 

In Table\,\ref{t:x-match_summary}, we summarise the results of the cross-match procedure. It is important to stress here that we used the APASS-PPMXL-{\em GALEX}  (APG, hereafter) cross-match
to characterise and identify most of the white dwarfs described in this paper. We used {\em WISE} photometry to probe for infrared excess, e.g. the binary systems discussed in Section\,\ref{chap6.4}. 

To visualise the cross-match procedure,  we display in Fig.\,\ref{f:cross_match} the identification of a newly confirmed white dwarf,  J0515+5021 (GD\,289). We note the star's significant proper motion across different epochs when the considered set of images was taken.
\begin{table}
\centering
\caption{Results of the all-sky cross-match.\label{t:x-match_summary}}
  \begin{tabular}{@{}lr@{}}
  \hline
Survey &  no. of objects \\ 
 &    \\ 
  \hline
APASS DR9 & 61,176,401\\
APASS DR9 unique & 58,756,366\\
APASS + {\em GALEX} (FUV and NUV) & 19,037,824 \\
APASS + {\em GALEX} NUV unique & 15,558,208 \\
APASS + PPMXL & 44,290,772\\
APASS + 2MASS & 42,901,844\\
APASS + {\em WISE}     & 43,179,834 \\
\hline
\end{tabular}
\end{table}
\begin{figure}
\centering
\includegraphics[width=\linewidth]{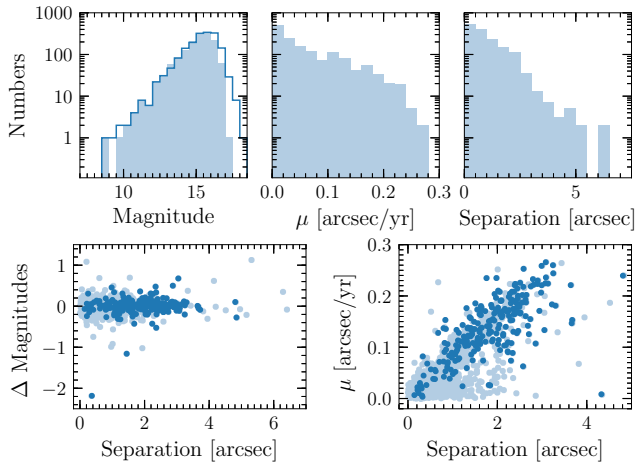}
\caption{Assessment of the APG cross-match for the GD sample. ({\em Top}): solid histograms represent the APASS $V$-band magnitudes, proper motions, and angular separations between APASS coordinates and those available in VizieR. The step histograms represent the $V$-band magnitudes 
available on VizieR . ({\em Bottom}): on the left, the differences between APASS magnitudes and those from the literature are plotted against the angular separation; on the right, we show the correlation between PPMXL proper motions, plotted against the angular separations. We use dark points to represent known white dwarfs and light coloured points for non white dwarf sources.}
\label{f:comparison_gd}
\end{figure}
\begin{figure}
\centering
\includegraphics[width=\linewidth]{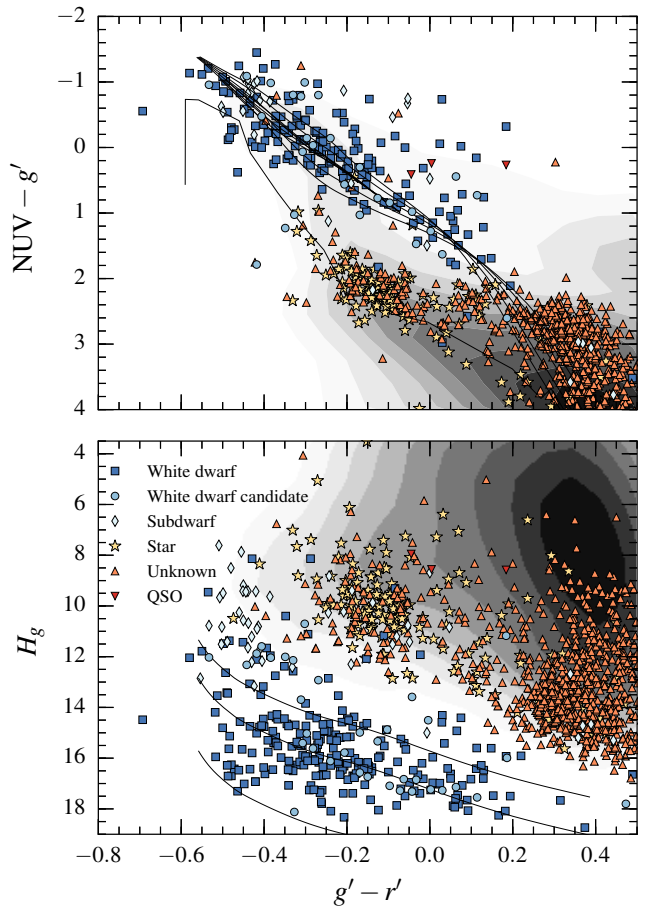}
\caption{Colour-colour (top) and reduced proper motion (bottom) diagrams of the GD sample. The dark coloured contours in both panels represent the colour distribution of the field population, as inferred from our all-sky cross-match. The object classification is based on the information available from SIMBAD (see text). ({\em Top}):  white dwarf cooling sequences ($\log{g} = 7.0$--$9.5$ in 0.5\,dex steps) and the main sequence are plotted for reference;  both have been  determined by convolving \citet{Koester10} synthetic spectra and \citet{Pickles98} spectra with the filter transmission profiles, respectively. ({\em Bottom}): solid curves representing the expected reduced proper motions for $\log{g} = 8$ disc white dwarfs are plotted at three distinct values of tangential velocities,  $V_{\rm T} = 20$, 40, and 150\,km/s. The curves follow the relation: $H_{g} = g' + 5 \times \log{V_{\rm T}} - 3.379$. }
\label{f:GD_pm}
\end{figure}
\section{Assessing the cross-match}
\label{chap3}
To verify the completeness of the APG cross-match, we compared the data to a set of catalogues from the literature. Our first comparison is done against the GD sample 
\citep[][]{Giclas80}  that contains blue stellar objects 
with proper motions of $\lesssim 0.20$\,arcsec/yr and includes 311 previously known white dwarfs. In this work, we
have confirmed 12 new GD stars as white dwarfs. 

During this operation, we visually inspected finding charts to confirm our re-identifications. We download cut-out images from the NASA SkyView service\footnote{\url{http://skyview.gsfc.nasa.gov/}},  i.e. blue and red frames of the STScI Digitized Sky Survey\footnote{\url{https://archive.stsci.edu/cgi-bin/dss_form}}, 2MASS, and {\it GALEX}, and for known white dwarfs we used finding charts available at the DENSE webpages, or those made available by Tom Marsh\footnote{\url{http://deneb.astro.warwick.ac.uk/phsaap/wdcharts/}}.

We make all the photometry and proper motions for the re-identified stars available on VizieR 
\citep[hosted by the Strasbourg astronomical Data Center, CDS;][]
{Wegner00}. Table~\ref{t:schema} describes the columns that are included in the online tables.

\subsection{The GD sample}
\label{chap3.1}

The catalogue of blue stars from the Lowell Observatory proper motion survey contains 1710 unique objects. For this test we used the list of coordinates and magnitudes that is available on VizieR (catalogue ID, II/299); these quantities are an updated collection of data from other surveys and not just from the original work. Through our cross-match described above, we successfully retrieved APASS photometry, PPMXL proper motions, and {\em GALEX} photometry  for 1479, 1455, and 1175 stars, respectively. Of these stars, we re-identify APASS photometry for 257 white dwarfs. In Fig. \ref{f:comparison_gd}, the distribution of these magnitudes is compared to those from the GD catalogue as available on VizieR. In Fig. \ref{f:comparison_gd}, we also display the distribution of PPMXL proper motions ($\mu$) and angular separations between the APASS coordinates and those listed in VizieR. In the bottom left panel of Fig.\,\ref{f:comparison_gd},  we display the difference between the APASS\,$V$ magnitudes plotted against the angular separations (APASS -- VizieR coordinates). We measure a scatter of $\pm 0.15$\,mag around the zero.  Five white dwarfs are at more than 5-$\sigma$ from the zero: GD\,123, which has an M-dwarf companion; GD\,167, for which the $V$-band photometry recorded in VizieR is $\approx 2$\,mag brighter than those measured in other bands available on SIMBAD (also hosted by CDS); GD\,230, GD\,272, and GD 546, for which we suspect the APASS photometry to be blended with nearby objects. In the bottom right panel of Fig.\,\ref{f:comparison_gd}, we display the correlation between PPMXL proper motions and angular separations. The majority of GD stars follow a linear trend,  which is more evident for the known white dwarfs in the sample. After inspecting finding charts for these stars, we note that most of the scatter appears to be due to the limited accuracy of the reference positions that are listed in VizieR. The most discrepant white dwarfs, GD\,230 (with an angular separation of 9-arcsec, off scale in Fig.\,\ref{f:comparison_gd}), is likely blended with nearby stars.

We inspected the literature on GD stars available through SIMBAD and divided them into six groups: white dwarfs, white dwarf candidates, subdwarfs, stars, unknown objects, and QSOs. More than half of the GD stars have very little information available (i.e. no classification in the references) or no SIMBAD entry at all. In Fig.\,\ref{f:GD_pm}, we display the colour-colour and reduced proper motion  diagrams of stars with APASS-PPMXL-{\em GALEX} data (the reduced proper motion is defined as $H_{g} = g' + 5 + 5 \times \log{\mu}$, with the proper motion, $\mu$, given in arcsec/yr). We note that confirmed white dwarfs and hot subdwarfs mostly separate from the main group of other stellar objects. Based on the colours, reduced-proper motions, and the information available in SIMBAD, we identify 46 candidate white dwarfs, of which we observed 13 spectroscopically. In the GD sample, there are about 1100 objects with little or no information on SIMBAD, which have colours and proper motions that  compatible with those of main sequence stars.  

\subsection{White dwarf samples}
 As seen from the previous test, the re-identification of known objects may present 
some issues due to the poor accuracy of reference positions. This can be accentuated for
bright white dwarfs, because their coordinates were traditionally recorded in equinox B1950.0 with an accuracy of the order of 1\,arcmin \citep[e.g.][]{McCook99} and possess on average large proper motions. With the advance of technology, 
higher accuracy is required, and there is an ongoing effort for updating the
astrometry and proper motions of known white dwarfs,
e.g. on SIMBAD and the Montr\'eal White Dwarf Database \citep[MWDD;][]{Dufour16}.

To assess the completeness of the APG cross-match with respect to 
well studied white dwarfs, we took into consideration five spectroscopic catalogues that 
include about 2/3 of the spectroscopically confirmed white dwarfs within the APASS magnitude range.  These are: i) the 25\,pc sample \citep[analysed by][]{Sion14}; ii) hydrogen (DA) and and helium (DB/DBA) white dwarfs from the European Southern Observatory  (ESO) supernovae Ia progenitor survey \citep[SPY;][respectively]{Koester09,Voss07}; iii) DA and DB white dwarfs analysed by the Montr\'eal group \citep[to be distinguished from the MWDD, which contains
a large fraction of the spectroscopically confirmed white dwarfs;][]{Gianninas10, Bergeron11}; iv) a bright selection ($g \leq 17$) of white dwarf plus main sequence binaries (WD+MS) from SDSS \citep[][]{Rebassa12}.  Covering most of the sky with Galactic latitudes of $|b|>30$\,deg, these catalogues allow for a semi-independent assessment of the photometric,  proper motion, and on-sky completeness of our APG dataset. We note that
for each catalogue we used the coordinates available on-line through VizieR, or SIMBAD. 
Given that the accuracy of positions may be different in each catalogue, 
this can lead to possible mismatches
and/or to non-identifications even for objects that are in more than one of the considered catalogues. When assessing the global-completeness, 
we took in account the unique objects across the catalogues.
\subsubsection*{The local 25\,pc sample}
\begin{figure}
\centering
\includegraphics[width=\linewidth]{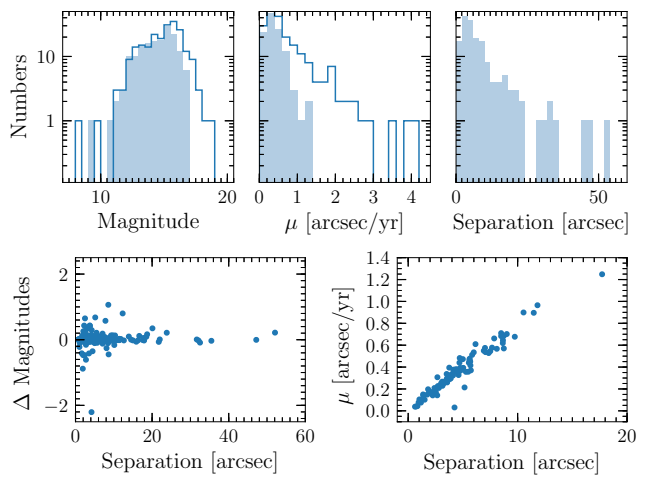}
\caption{Assessment of the APG cross-match for white dwarfs in the 25\,pc sample. Refer to Fig.\,\ref{f:comparison_gd} for a description of each panel.}
\label{f:comparison_sion14}
\end{figure}

For this comparison, we use the list of \citet{Sion14}, which contains 223 white dwarfs. We identify APASS matches for 169 stars. The unmatched objects include Sirius\,B or Procyon\,B, which are outshined by their bright main sequence companions. In the top and bottom panels of Fig.\,\ref{f:comparison_sion14}, we display the comparison between APASS\,$V$ photometry and data available on SIMBAD for the \citet{Sion14} white dwarfs in the same band. For this sample, we measure a mean difference of $\pm 0.35$\,mag. Most of the outliers are due to contamination from nearby sources. We note that our cross-match becomes increasingly incomplete below 16.5\,mag, and we do not retrieve any of the white dwarfs fainter than 17\,mag (5 per cent of the total). The brightest white dwarf we re-identify is 40\,Eri\,B (APASS~$V = 9.3$).  

The completeness of the cross-match with the \citet{Sion14} sample is reduced to 116 white dwarfs when we consider PPMXL proper motions. The highest proper motion we retrieve is 1.25\,arcsec/yr for WD\,2248+293. From Fig.\,\ref{f:comparison_sion14}, we note that we are about 60 per cent complete for white dwarfs with proper motions $\leq 1$ arcsec/yr.
In the top right panel of Fig.\,\ref{f:comparison_sion14}, we present the distribution of angular separations between APASS and SIMBAD coordinates for the \citet{Sion14} white dwarfs. 40\,Eri\,B  is the star with the largest angular separation (52-arcsec). We do not plot WD\,0727+482A (with an angular separation of 90-arcsec, off-scale in Fig.\,\ref{f:comparison_sion14}), for which SIMBAD lists the coordinates of the companion.
 
In the bottom right panel of Fig.\,\ref{f:comparison_sion14}, we display the correlation between angular separations (APASS -- SIMBAD coordinates) and the proper motions. First, we note that the correlation follows closely a linear trend. Second, we fail to retrieve proper motions for stars with separations that are larger than 20 arcsec. Most of these objects move faster than 1.5 arcsec/yr. The lowest point not aligning with the others is WD\,0108+277, the astrometry of which may be affected by the presence of a nearby background source \citep[][]{Farihi05, Holberg16}.

\subsubsection*{White dwarfs from the SPY project}
\begin{figure}
\centering
\includegraphics[width=\linewidth]{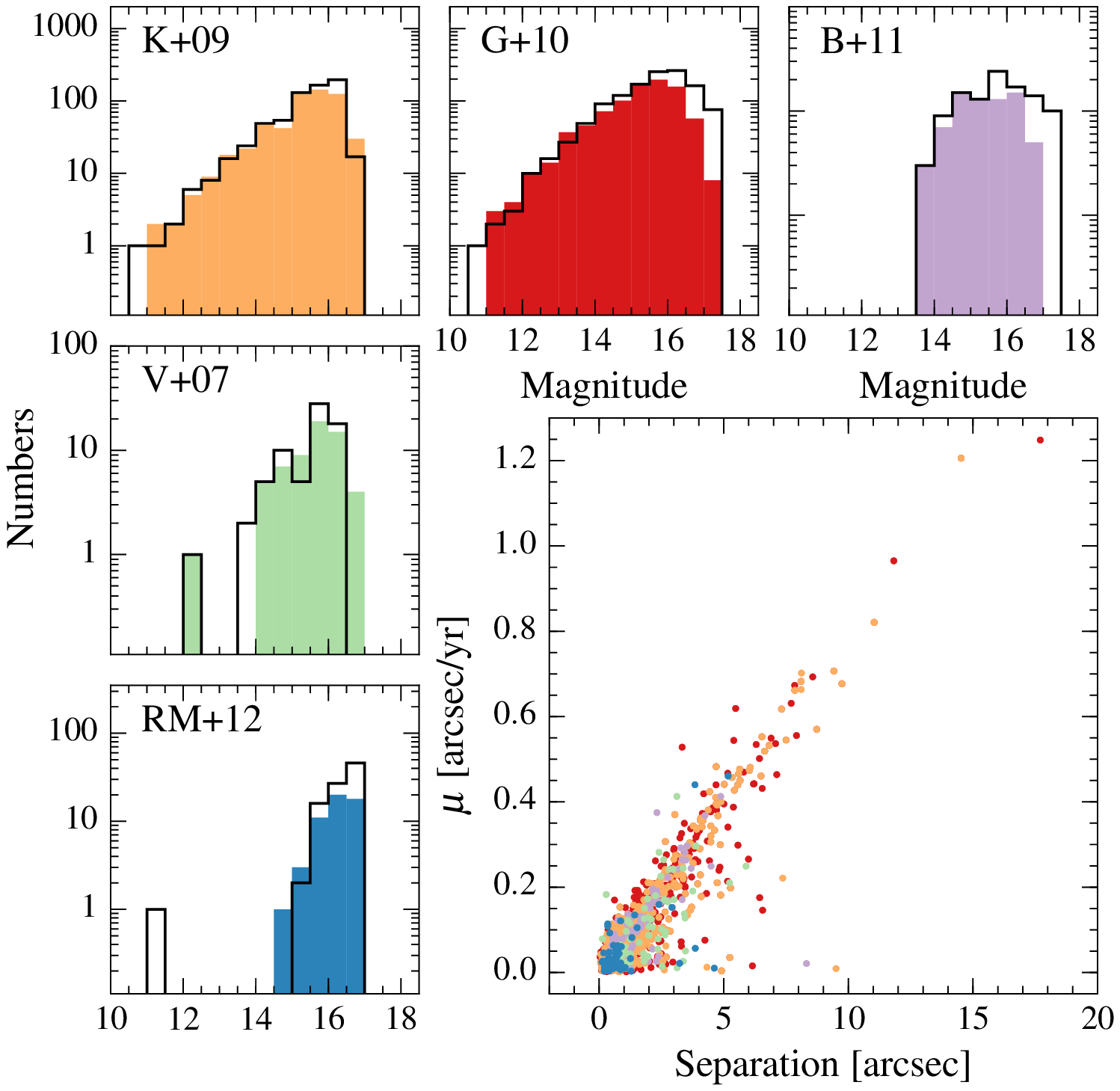}
\caption{Validation of the APASS-PPMXL cross-match. The coloured histogram bars represent APASS magnitude distributions ($V$ or $g'$, depending on the sub-sample considered) for the five white dwarf catalogues we used as reference: K+09: \citet{Koester09}; V+07: \citet{Voss07}; G+10: \citet{Gianninas10};   B+11: \citet{Bergeron11}; RM+12: \citet{Rebassa12}. The black outline-histograms represent the magnitude distributions of the known white dwarfs as retrieved from the literature. In the bottom-right panel, we plot the PPMXL proper motions against the angular separations between APASS coordinates and the positions available on VizieR/SIMBAD. Symbols are colour-coded with the same colours used of the histogram bars.} 
\label{f:comparison_samples}
\end{figure}
SPY was an ESO spectroscopic survey that targeted bright white dwarfs ($B \leq 16.5$) south of $\delta = +25$\,deg \citep[][]{Napiwotzki03}. We tested the completeness of APASS on the DA and DB/DBA white dwarf catalogues published by \citet{Koester09} and \citet{Voss07}, respectively. We re-identified 575 of the 669 DA white dwarfs, and 61 of the 69 DB/DBA white dwarfs from SPY. The comparison between APASS magnitudes and those available in the literature is shown in the left panels of Fig.\,\ref{f:comparison_samples}. 

We retrieved PPMXL proper motions for 547 DA and 59 DB/DBA white dwarfs. On average, proper motions of these stars are smaller in comparison with those of white dwarfs in the local sample, therefore we have an higher success rate in re-identifying these stars through our selection method. In the bottom-right corner of Fig.\,\ref{f:comparison_samples}, we display the correlation between PPMXL proper motions and angular separations between APASS and Vizier coordinates for the SPY sample. In the same panel, we plot also white dwarfs from the other samples discussed next. Proper motion and angular separations correlate well as found above for the \citet{Sion14} sample. Just a few objects (e.g. WD\,1334-678, WD\,1943+163, WD\,2029+183, and WD2157+161), which are blended with nearby stars may have inaccurate positions and/or proper motions.
\begin{table*}
\centering
\caption{Summary of the comparison between the APG cross-match with stars from the literature. \label{t:x-match_literature}}
  \begin{tabular}{@{}lccccl@{}}
  \hline
Catalogue &  no. of objects & APASS & PPMXL & {\em GALEX}\,NUV & reference \\ 
\hline
GD sample (all objects) &  1710  & 1479 & 1455 & 1175&\citet{Giclas80}     \\ 
GD sample (confirmed white dwarfs) &  323  & 260 & 246 & 174&     \\ 
25 pc & 223 & 169 & 115& 40&\citet{Sion14}     \\ 
SPY DA & 669 & 575 & 547 & 394&\citet{Koester09}     \\ 
SPY DB/DBA & 69 & 62 &59& 37 &\citet{Voss07}     \\ 
Montr\'eal DA & 1265 &889 & 822 & 546&\citet{Gianninas10}    \\ 
Montr\'eal DB/DBA & 108 & 72 & 70 & 48&\citet{Bergeron11}     \\ 
WD+MS & 92& 55 & 54 & 39&\citet{Rebassa12}     \\ 
\hline
Total white dwarfs & 1931 & 1388 & 1271 & 840 &    \\
\end{tabular}
\end{table*}
\begin{figure}
\centering
\includegraphics[width=\linewidth]{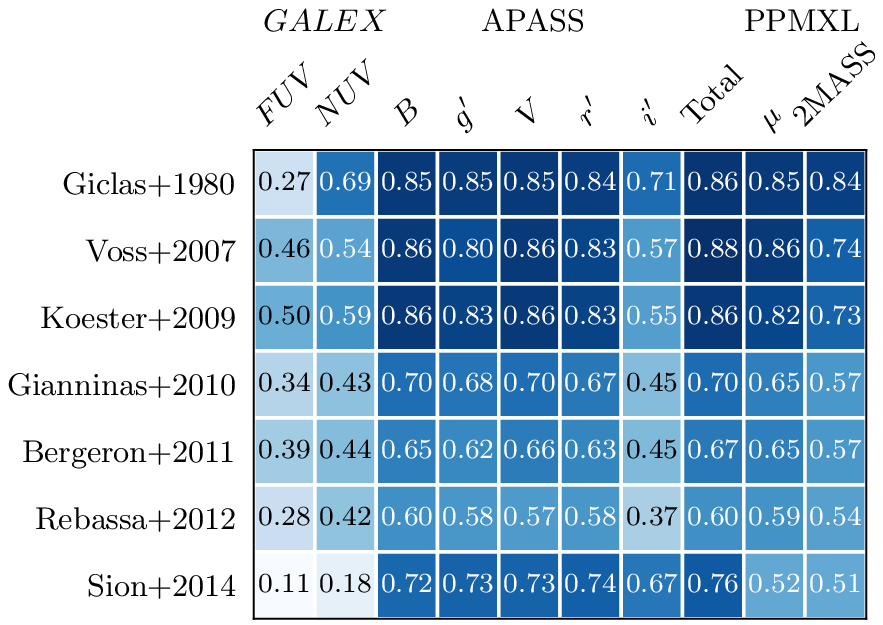}
\caption{Completeness of the dataset, expressed as fraction of the number of stars retrieved with respect to the total number of objects in the corresponding sample.}
\label{f:completeness}
\end{figure}
\begin{figure*}
\centering
\includegraphics[width=\linewidth]{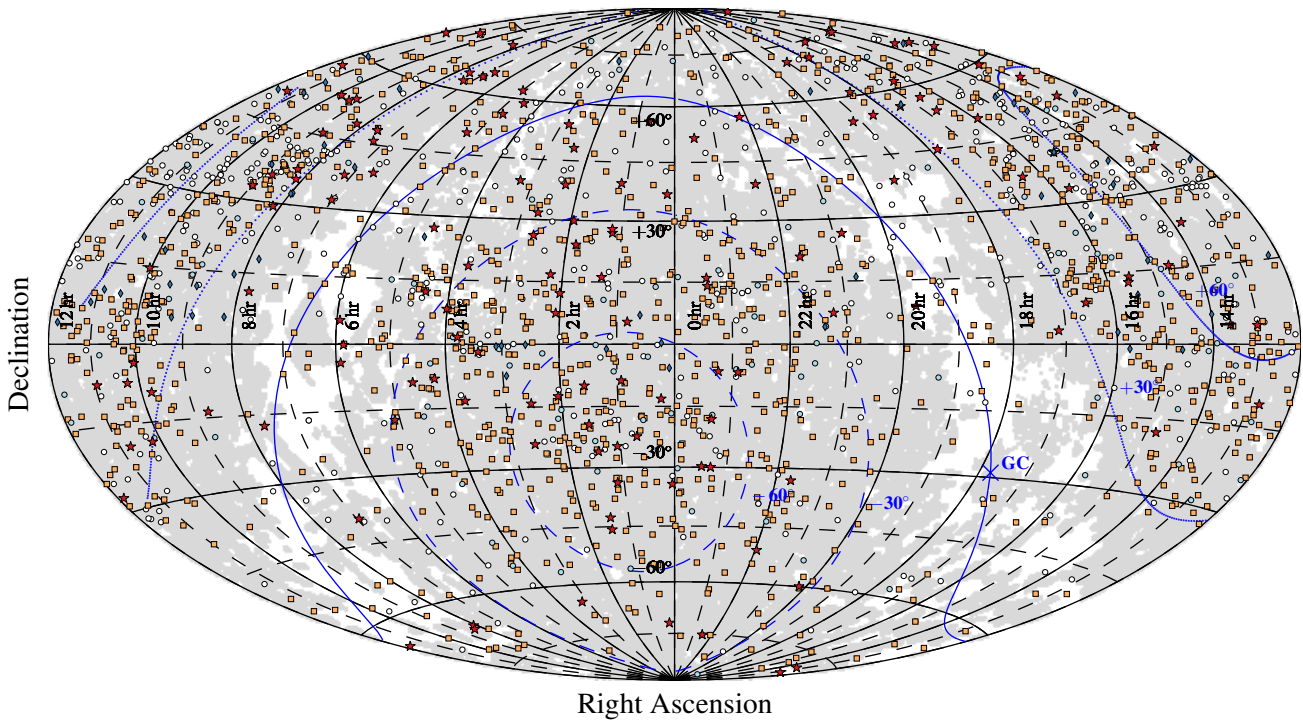}
\caption{Spatial distribution in equatorial coordinates of the known white dwarfs from the literature discussed here. Coloured symbols are used for the objects we re-identify in our multi-band photometric dataset (squares: DA white dwarfs; circles: DB/DBA white dwarfs; diamonds: WD+MS; stars: 25\,pc sample). White symbols correspond to white dwarfs with no APASS cross-match. The light grey-shaded area delimits the sky coverage of {\em GALEX}. 
Iso-Galactic latitude curves are overplotted in blue.}
\label{f:samples_distribution}
\end{figure*}

\begin{figure*}
\centering
\includegraphics[width=0.85\linewidth]{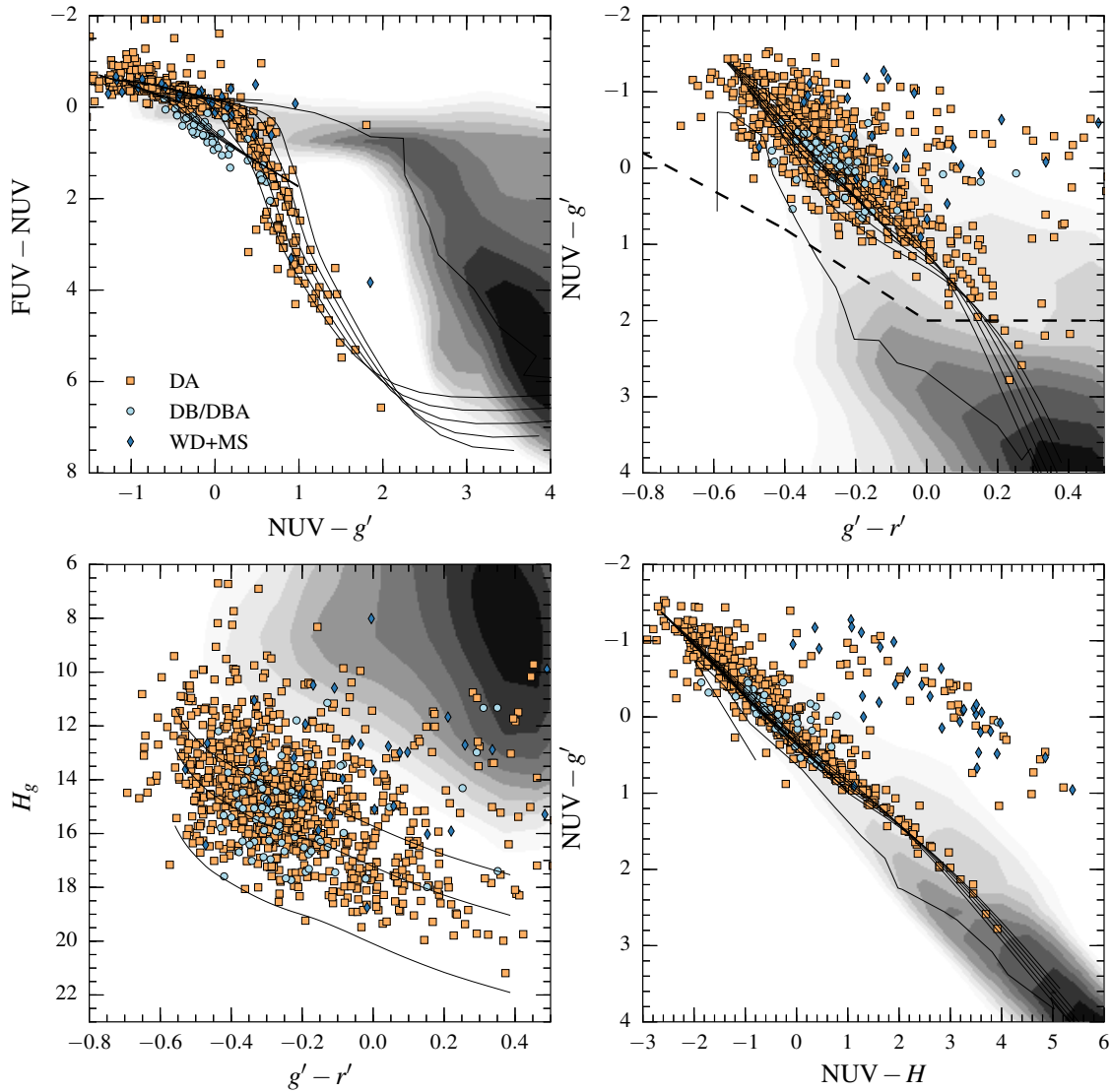}
\caption{Colour-colour and reduced proper motion diagrams, obtained by combining {\em GALEX}, APASS, and 2MASS photometry. The number of objects plotted in each panel depends on the availability of the data (refer to Table\,\ref{t:x-match_literature} and Fig.\,\ref{f:completeness}). The dark coloured contours in both panels represent the colour distribution  of the field population, as inferred from our all-sky cross-match. Symbols used to distinguish distinct classes of white dwarfs are listed in  the top-left panel. The dashed curves in the top-right panel delimit the colour space where most known white dwarfs are found. White dwarf cooling sequences ($\log{g} =7.5$--$9.5$ in 0.5\,dex steps) and main sequence are plotted for reference in the three colour-colour diagrams. The expected proper motion curves for $\log{g}= 8$ Galactic disc white dwarfs are plotted in the bottom-left diagram. Also refer to the caption of  Fig.\,\ref{f:GD_pm} for further details. }
\label{f:samples_ccd_selection}
\end{figure*}
\begin{figure}
\centering
\includegraphics[width=\linewidth]{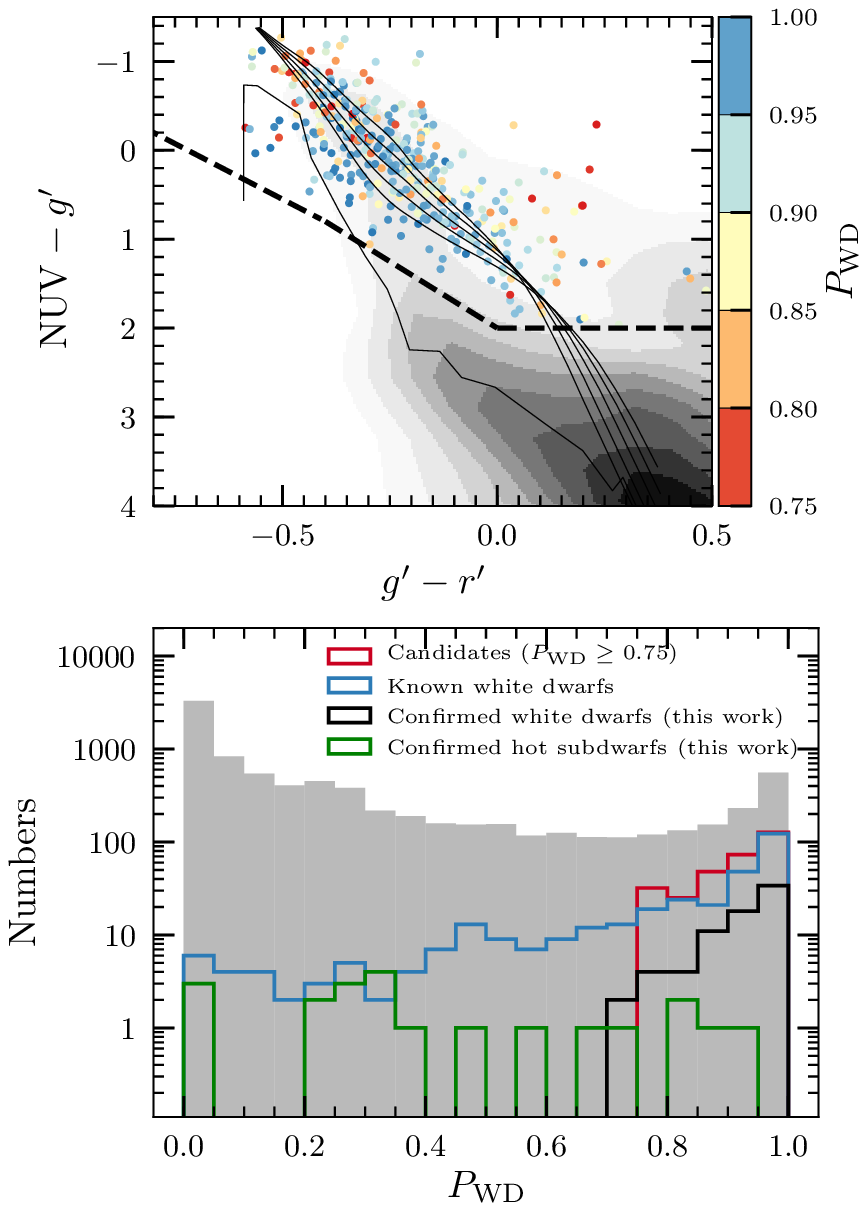}
\caption{Selection of white dwarf candidates. ({\em Top panel}): colour-colour diagram showing the colour-cut (dashed lines) applied to select white dwarf candidates. The side-bar codes the assigned probability of being white dwarf, $P_{\rm WD}$, based on the reduced proper motion information. ({\em Bottom panel}): histogram distribution, on a logarithmic scale, of 1.5 million objects with $g'-r'\leq 0.5$ and ${\rm NUV} - g' \leq 4$, which we bin in ranges of $P_{\rm WD}$. We overlay the histogram distribution of  $P_{\rm WD}$ for known white dwarfs, candidate white dwarfs, and the white dwarfs and hot subdwarfs that we confirmed spectroscopically. }
\label{f:cut_pwd}
\end{figure}
\subsubsection*{DA and DB white dwarfs analysed by the Montr\'eal group}
These two white dwarf samples overlap to some degree with SPY, and consist of 1265 DA \citep[][]{Gianninas10} and 108 DB/DBA white dwarfs \citep[][]{Bergeron11}. The relevant differences with the SPY sample are their spatial distribution that extends to the northern hemisphere and a fainter limiting magnitude ($\approx 17.5$\,mag). We identified APASS photometry for 889 DA and 72 DB/DBA white dwarfs. The magnitude distributions of both samples are displayed in the top-right panels of Fig\,\ref{f:comparison_samples}. We note that the completeness starts decreasing for stars that are fainter than 16.5\,mag. In the bottom right panel of Fig\,\ref{f:comparison_samples}, we find a good correlation between proper motions and angular separations also for these stars; like before, a few discrepant objects (e.g. WD\,0927$-$173 or WD\,1545+244) are close to nearby sources that cause less accurate measures of proper motions or positions.
\subsubsection*{White dwarf plus main sequence star binaries}
\citet{Rebassa12} have produced  the so far largest catalogue of WD+MS binaries\footnote{\url{http://www.sdss-WDMS.org/}} from SDSS, which has been recently updated to the data releases 9 and 12 \citep[][]{Rebassa16}.  We selected a sample of 92 binaries with $g' \leq 17$, re-identifying APASS and PPMXL data for 55 and 54 of them, respectively. The magnitude distribution of the sample is displayed  in the bottom-left corner of Fig.\,\ref{f:comparison_samples}, while in the the bottom-right panel we add the corresponding data points to the comparison between PPMXL proper motions and coordinate separations.

\subsection{Completeness}
Table\,\ref{t:x-match_literature} summarises the tests of our combined 
APG catalogues. The completeness of the cross-match depends on the intrinsic properties of the different test cases, e.g. the magnitude limits and accuracy of positions. In total, we have retrieved APASS photometry for 1388 white dwarfs out of a sample of 1931 stars (72 per cent). 
We note that that the median magnitude of the white dwarfs for which we do not retrieve APASS data is $V \approx 16.5$, below which we are about 80 per cent complete. The completeness with respect to PPMXL and {\em GALEX} photometry is 66 and 43 per cent, respectively. A more detailed breakdown  into survey passbands is displayed in Fig.\,\ref{f:completeness}, for the spectroscopic catalogues considered earlier.

In Fig.\,\ref{f:samples_distribution}, we display the spatial distributions of the white dwarfs discussed in the previous sections. We checked for spatial clustering of the white dwarfs we could not re-identify. Most of the unidentified white dwarfs that were not recovered in our merged APG catalogue are in the Montr\'eal sample and are located along a strip stretching from $\alpha = +0$--$5$\,hr and $\delta = -30,\,+15$\,deg to  $\alpha = 8$--15\,hr and $\delta = +30,\,+40$\,deg. Some of them are from the Kiso survey \citep[][and references therein]{Limoges10},  which overlaps with the fainter end of APASS. We also note the traditional deficit of white dwarfs at low Galactic latitudes that is common to all the known samples of white dwarfs. At low latitudes, there is also no {\em GALEX} photometry, mainly due to the satellite avoiding bright stars and crowded areas in the Galactic plane (note the white areas in Fig.\,\ref{f:samples_distribution}). 

Finally, we confirmed our results by checking all the remaining white dwarfs (where we define white dwarfs as objects with $\log{g}\geq6$) in the MWDD and SDSS \citep{Kleinman13,GentileFusillo15,Kepler15,Kepler16}, that is $\approx 1000$ objects brighter than 17~mag.  The overall completeness decreases to 62 per cent (1864 out of 2990 white dwarfs),  
given that most of these additional stars are in the range of $g=16-17$.

\subsection{Candidate selection}
\label{chap3.4}
Having characterised the photometry of known white dwarfs,
we can use our merged APG catalogue to identify new white dwarf candidates. In Fig.\,\ref{f:samples_ccd_selection}, we display three colour combinations and a reduced proper motion diagram for the known white dwarfs. We  corrected {\em GALEX} photometry for non-linearity effects, using the results of  \citet{Camarota14}. In the top-left panel of 
Fig.\,\ref{f:samples_ccd_selection}, DA white dwarfs are clearly distinct from main sequence stars when they reach $T_{\rm eff} < 12\,000$\,K $({\rm FUV} - {\rm NUV} > 1.3 ,\,{\rm NUV} - g' > 0.5)$.
In the top-right panel of Fig.\,\ref{f:samples_ccd_selection}, the $({\rm NUV} - g',\,g' - r')$ combination of colours is a proxy for the more commonly used SDSS $(u-g,\,g-r)$ diagram. The {\em GALEX}~NUV band samples a spectral range that is further away from the Balmer jump, resulting in a less stretched distribution of colours. We note a fairly good agreement between observed colours and the white dwarf cooling tracks, with an amount of scatter that is mostly due to the larger uncertainties of the APASS\,$r'$ magnitudes. It is worth noting the larger offset of some of the stars towards redder colours, which is a clear signature of the presence of late-type companions. The colour excess due to late-type companions becomes markedly evident when using 2MASS magnitudes, i.e. in the $({\rm NUV} - g',\,{\rm NUV} - H)$ panel of Fig.\,\ref{f:samples_ccd_selection}. Finally, the reduced proper motion diagram in the bottom left corner of this figure confirms the bulk of the re-identified stars as nearby, fast moving objects. 

To select white dwarf candidates, we applied an (${\rm NUV} - g'$, $g' - r'$) cut based on the colours of the re-identified stars (Fig.\,\ref{f:samples_ccd_selection}). This colour cut reduces the contamination from main sequence stars that may share a similar colour space with white dwarfs of $T_{\rm eff} \lesssim 7000$\,K.  We then defined a selection criterion following the work of \citet{GentileFusillo15} on SDSS, which establishes a probability for an object to be a white dwarf. This probability,  $P_{\rm WD}$,  is inferred from the colours and the reduced proper motion of known white dwarfs in the ($H_{g}$, $g' - r'$) plane. Given the larger uncertainties of PPMXL proper motions with respect to those from SDSS,  we only define objects with $\delta \mu/\mu \leq 0.3$ and $P_{\rm WD} > 0.75$ to be good white dwarf candidates. Below this value, we found that the APASS-PPMXL probabilities disagree by more than $\pm\,25$ per cent with  those measured by \citet[][]{GentileFusillo15} for a sample of $\approx 300$ stars in common. 

We identified 8000 objects within our colour-cut, of which $\simeq1200$ have $P_{\rm WD} > 0.75$.  Given the strict $P_{\rm WD}$ requirement, not all the known white dwarf fall within this selection, mostly because PPMXL proper motion uncertainties weight down their probabilities. In this
group of objects, we identified  391 that had not yet been spectroscopically confirmed as white dwarfs, ten of which are GD stars. We observed 77 high priority objects, 
confirming 71 white dwarfs and six hot subdwarfs.
Due to scheduling constraints we also included in our target list objects 
that fall outside the colour-cut or with $P_{\rm WD} < 0.75$,  
achieving a lower rate of confirmed white dwarfs.

After our spectroscopic follow-up, we are left with  
305 high-confidence white dwarf candidates with $P_{\rm WD} > 0.75$ and 33 GD stars
that do not have spectra or known spectral classification in the literature.  In Fig.\,\ref{f:cut_pwd}, we display the (${\rm NUV}-g',\,g'-r'$) colour-colour diagram of the candidate stars. In the lower panel of Fig.\,\ref{f:cut_pwd}, the histogram bars represent the $P_{\rm WD}$ of all objects with $g'-r'\leq 0.5$ and ${\rm NUV} - g' \leq 4$, i.e. about 1.5 million objects. We overlay step-histograms representing the $P_{\rm WD}$ of the known white dwarfs discussed in the previous sections, those for the 305 candidate white dwarfs, and those of white dwarfs and hot subdwarfs that we spectroscopically followed-up. 
 
The photometry and proper motions for the 305 white dwarf candidates are made available through VizieR
in the format as listed in Table\,\ref{t:schema}. The photometry for the 135 stars that we have followed up will also be hosted on VizieR, with the inclusion of an additional column, ``short name'', that we use through the paper to refer to specific stars.

\section{Spectroscopic follow-up}
\label{chap4}
We obtained low-resolution spectra for 135 stars covering the wavelength range from the Balmer jump to slightly beyond the H$\alpha$ (3700--7500\,\AA).  The observations were spread over 2.5 years (2014 Jun/Jul, 2015 Mar/May, and 2016 Nov/Dec), and the data were taken at three different telescopes: the 3.6-m New Technology Telescope (NTT) at the ESO La Silla observatory, the 2.5-m Isaac Newton Telescope (INT) at the Roque de Los Muchachos Observatory, and the 1.82-m Copernico Telescope at the Ekar Asiago Observatory. Our targets were drawn from the APG list of candidates, the GD sample, and some objects that have lower $P_{WD}$. 
We note that some of the targets (eight white dwarfs and ten subdwarfs; see next sections)
were previously known in the literature but with poorly accurate coordinates,
or that other research groups obtained classification during our follow-up operations.

At the NTT we used the ESO Faint Object Spectrograph and Camera 2 \citep[EFOSC2;][]{Buzzoni84} with the \#\,11 grism, a slit of 0.7\,arcsec, and binning 2x1 and 2x2, obtaining a resolution of 6.5--9.5\,\AA. The setup adopted at the INT produced a 3.5\,\AA\ resolution, mounting the Intermediate Dispersion Spectrograph (IDS) with the grating R400V a slit of 1\,arcsec and the blue-sensitive EEV10 CCD. For the Copernico setup,  we mounted the Asiago Faint Object Spectrograph and Camera (AFOSC) with the volume phase holographic grating, VPH7 \citep[][]{Zanutta14}, and a slit of 1.25\,arcsec, which deliver a resolution of 13\,\AA\ with a $2\times2$ pixels binning. The log of the observations is summarised in Table\,\ref{t:observing}. 

The observations were done under different sky illumination (grey/bright at the NTT, dark/bright at the INT, dark/grey at the Copernico), with the seeing typically $< 2.5$\,arcsec, variable weather conditions often with thin clouds and during non photometric nights (especially during the second run at NTT, the second run at INT when calima affecting the sky transparency occurred, and the November runs at Asiago that 
were affected by high humidity). We took 2--4 exposures per target, with exposure times  in the range of 300--1800\,sec.  On several occasions, we obtained multiple spectra of the same object across different runs to improve the total signal-to-noise ratio.

We took standard calibration frames (bias, dome flats) during each run, and arc lamps before and after the targets at the NTT and INT. We obtained a few arc lamps at beginning of the Asiago run (Ne and Hg+Cd separately, later combined during the data reduction), which were sufficient for an approximate wavelength calibration of the low resolution spectra. We also observed 2---3 spectrophotometric standards per night with a 5\,arcsec slit and we placed the slit at the parallactic angle for all the science and standard stars, in order to minimise flux-loss and to obtain a relative flux-calibration of the spectra, shown to be in good agreement with APASS photometry for most of the confirmed white dwarfs.  A few exceptions are the visually resolved WD+MS pairs, where we rotated the slit to include both the white dwarf and the late-type companion. 

We used standard reduction routines to subtract the bias, to apply the flat-field correction and wavelength calibration, to optimally-extract the 1D spectrum, and for the flux calibration. Specifically, we used the {\sc starlink} suite of programs including {\sc pamela}\footnote{\url{http://starlink. eao.hawaii.edu/starlink}}\citep[][]{Marsh89} and {\sc molly}\footnote{\url{http://www. warwick.ac.uk/go/trmarsh/software/}}.
\begin{table}
{\scriptsize
\centering
\caption{Summary of instruments and number of spectra for the follow-up discussed in this paper.\label{t:observing}}
  \begin{tabular}{@{}llcclr@{}}
  \hline
Telescope &  Setup & $\Delta \lambda$ & Slit & Date & $N$\\ 
 &   &  [\AA] & [''] & &   \\ 
  \hline
NTT   &  EFOSC2/\#\,11 & 9.5  & 0.70 &2014/06/13--15 & 48  \\
      &         & 6.5--9.5  & 0.70 & 2014/07/11--13 & 34   \\
INT   &  IDS/R400V  &  3.5  & 1 & 2015/03/03--06  & 46   \\
      &         &    3.5  & 1.0 &2015/05/12--18 & 38   \\
Copernico & AFOSC/VPH7 & 13 & 1.25 & 2016/11/05 & 2 \\ 
                           &  & 13 & 1.25 & 2016/12/26--30 & 7 \\ 
\hline
\end{tabular}}
\end{table}
\begin{table}
\centering
\caption{Summary of spectral classification.\label{t:typying}}
  \begin{tabular}{@{}ll@{}}
  \hline
Type &  Number \\  
  \hline
White dwarfs  (total) &  82  \\
     DA             &  60        \\
     DA+MS & 9         \\   
     DAH &   1       \\
     DAB &   1       \\
     DB/DBA &   4       \\
     DC &   7       \\
Hot subdwarfs/BHB  &  25  \\
Other stars  &  28  \\
\hline
\end{tabular}
\end{table}
\begin{figure*}
\centering
\includegraphics[width=\linewidth]{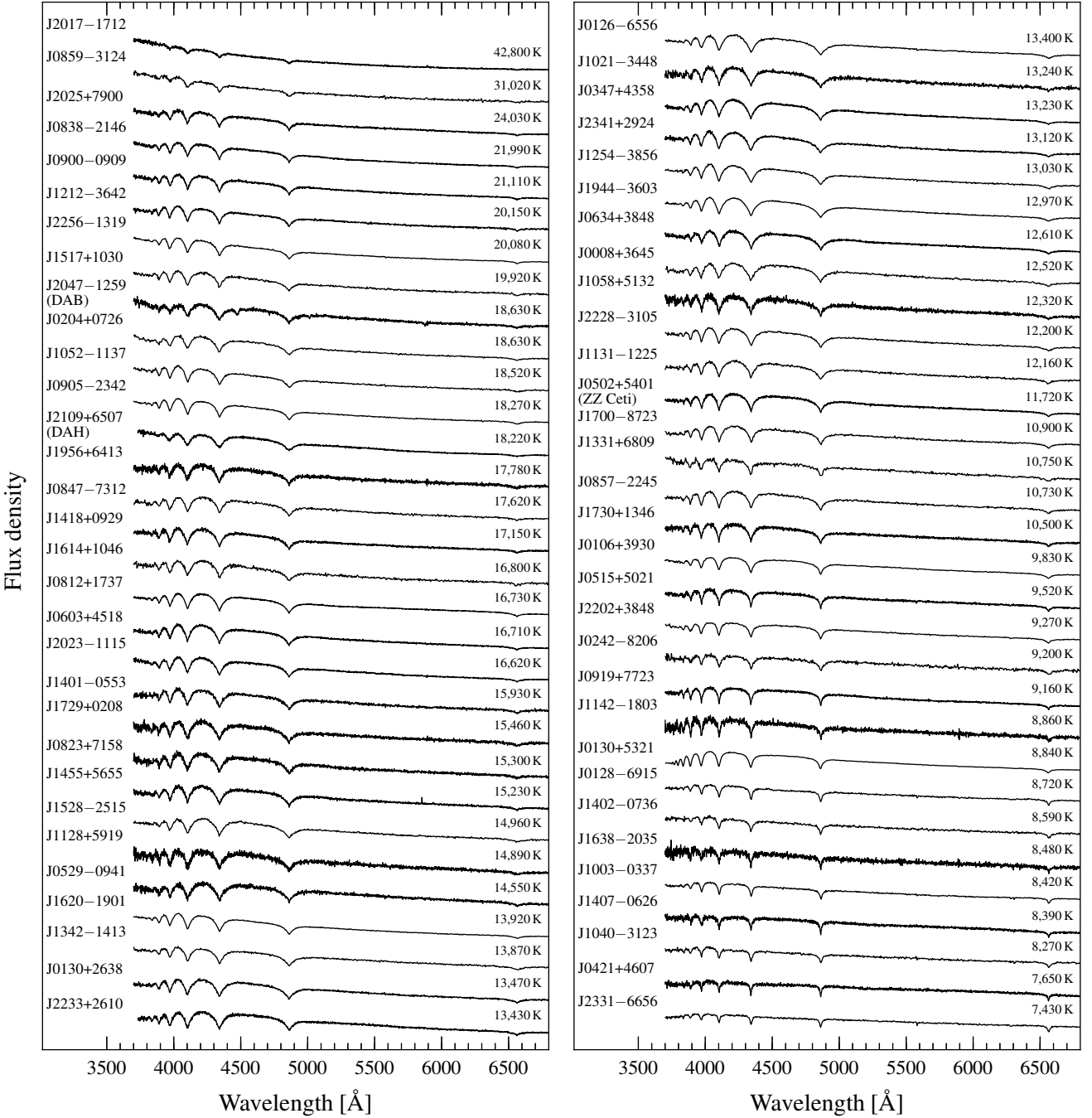}
\caption{Spectra of the single hydrogen-atmosphere (DA) white dwarfs, 
sorted by temperature. Where relevant, subtypes are included between brackets.}
\label{f:da}
\end{figure*}
\begin{figure}
\centering
\includegraphics[width=\linewidth]{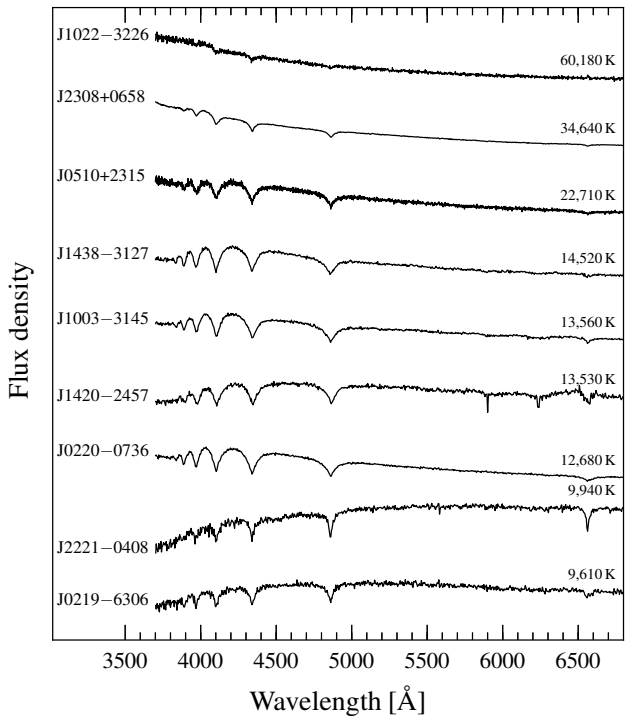}
\caption{DA+MS binaries sorted by temperature.}
\label{f:dams_spectra}
\end{figure}
\begin{figure}
\centering
\includegraphics[width=\linewidth]{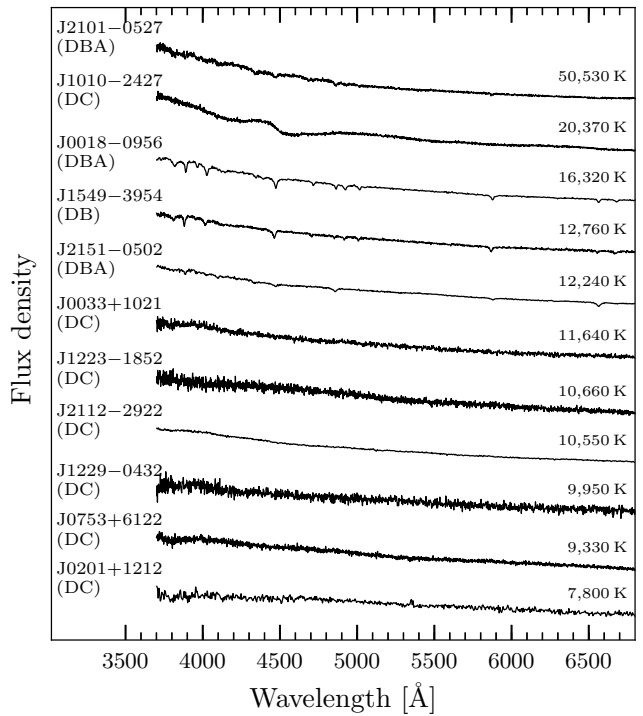}
\caption{DBA, and DC white dwarfs.}
\label{f:dbc}
\end{figure}
\begin{figure*}
\centering
\includegraphics[width=\linewidth]{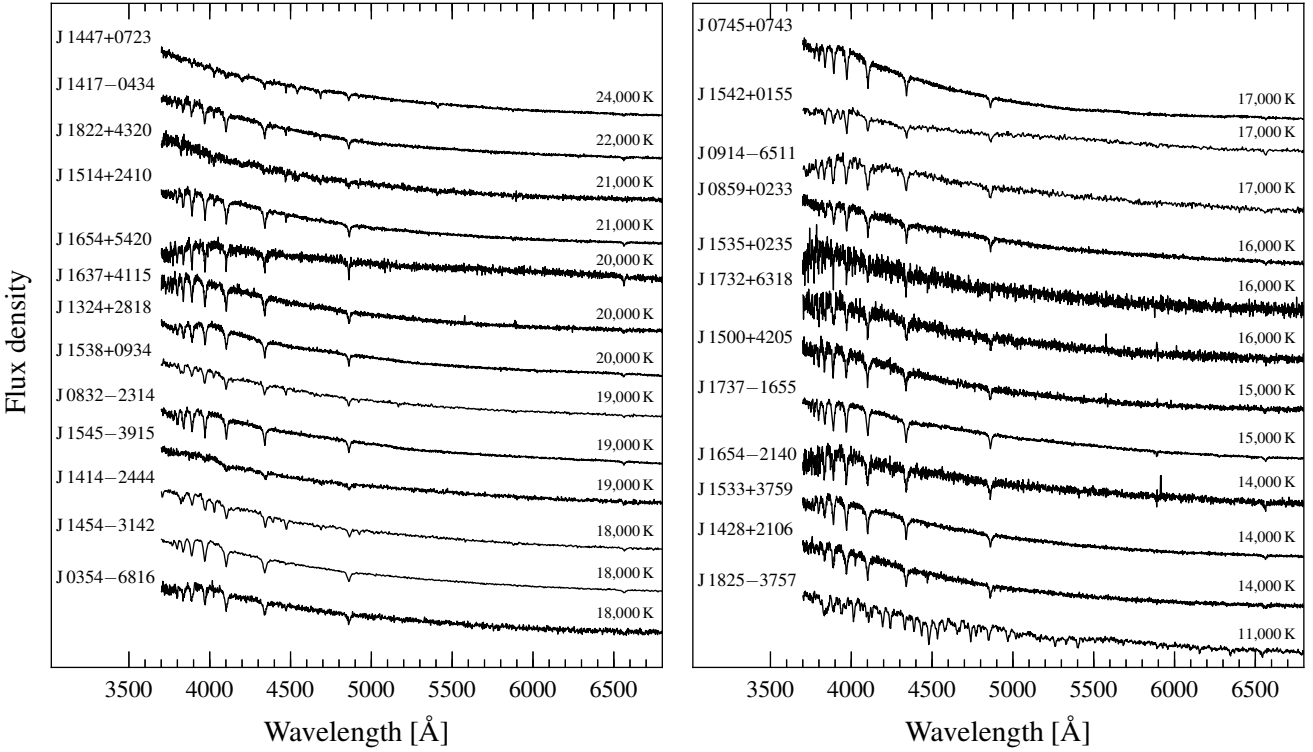}
\caption{Hot subdwarfs and BHB stars sorted by temperature.}
\label{f:sd}
\end{figure*}
\begin{figure*}
\centering
\includegraphics[width=\linewidth]{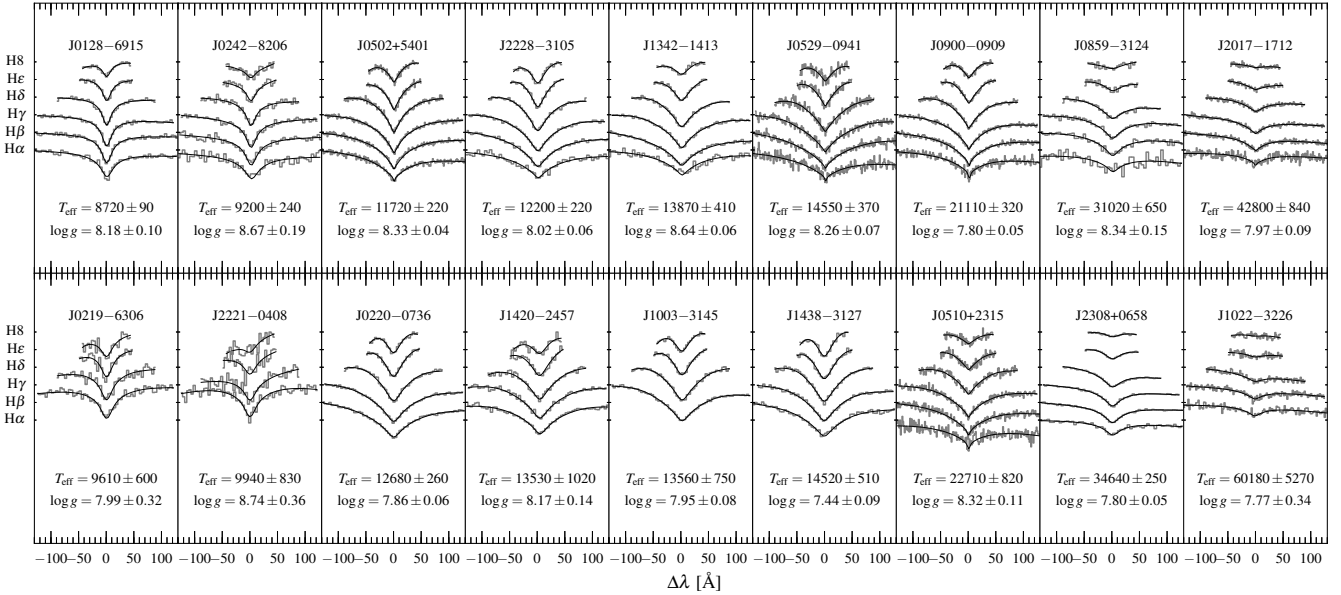}
\caption{Best fits of the Balmer lines for selected single DA white dwarfs (top row) and those in DA+MS pairs (bottom row). Synthetic spectra (black) are overlaid onto observed spectra (grey).}
\label{f:da_lines}
\end{figure*}
\begin{figure}
\centering
\includegraphics[width=\linewidth]{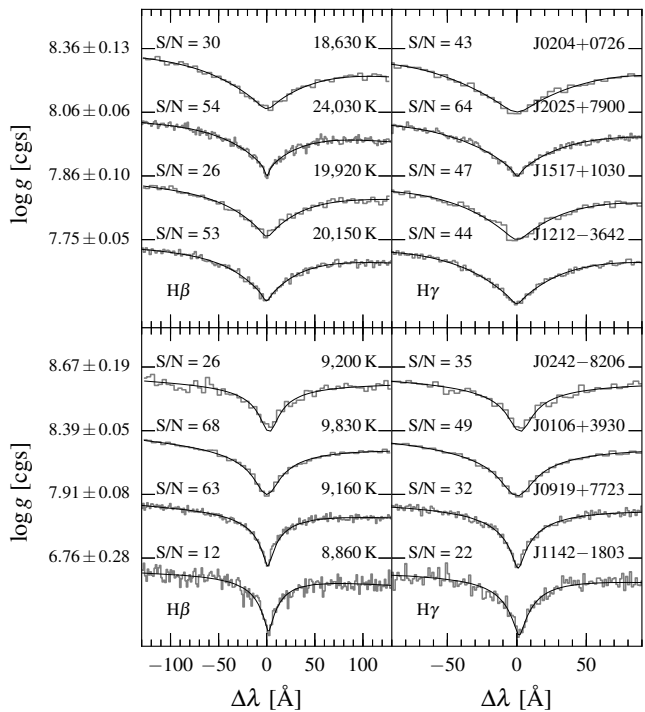}
\caption{Best fits of H$\beta$ and H$\gamma$ lines for selected objects with $T_{\rm eff} \approx 20\,000\,K$ (top row), and $\approx 10\,000\,K$ (bottom row). Objects are plotted in order of decreasing $\log{g}$. Synthetic spectra (black) are overlaid on to observed spectral lines (grey).}
\label{f:da_logg}
\end{figure}
\begin{figure}
\centering
\includegraphics[width=\linewidth]{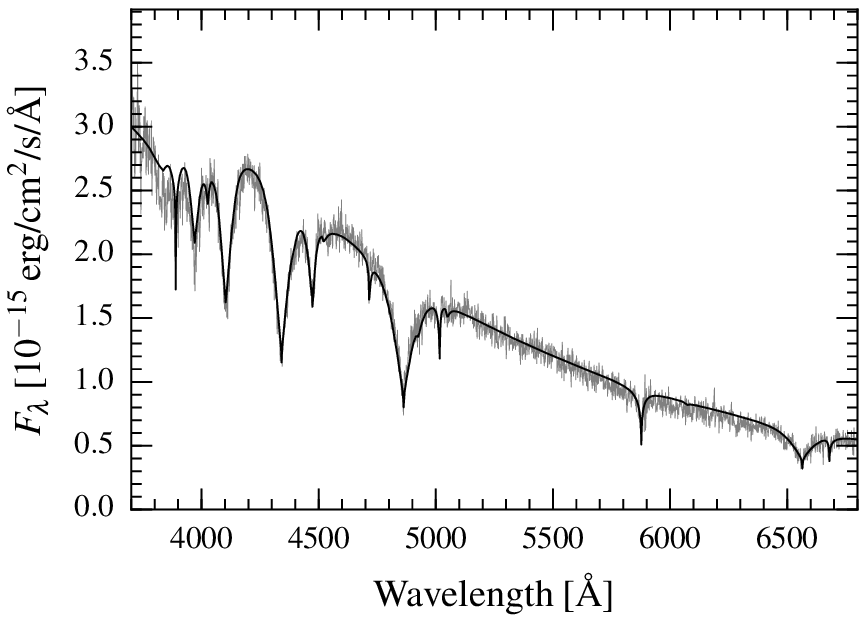}
\caption{Observed spectrum (grey) of the DAB white dwarf, J2047$-$1259, and the best-fit model spectrum (black) with $T_{\rm eff} = 18\,630 \pm 360$, $\log{g} = 8.37 \pm 0.6$, and [H/He]$ = -1$.}
\label{f:dab}
\end{figure}
\subsection{Spectral classification}
From a visual inspection of the multiple observations, we selected the best spectra for each star that we used for a quantitative classification. These spectra have a typical signal-to-noise ratio larger than 20. We confirmed 82 white dwarfs (eight of which were previously known) and 25 hot subdwarfs/blue horizontal branch (BHB) stars (ten of which were previously known). We classify the remaining 28 objects as A/F/G dwarfs, some of which have apparently large proper motions and colours that overlap with those of cool white dwarfs (see Fig.\,\ref{f:follow-up_ccd}). 

The breakdown of the sample by sub-type is summarised in Table\,\ref{t:typying}. We identify 71 hydrogen atmosphere white dwarfs; in Fig.\,\ref{f:da}, we show 60 single objects (DA), one magnetic white dwarf with Zeeman splitting of the Balmer lines (DAH), and another that presents also helium lines (DAB). We classify nine white dwarfs as binaries with a late type companion (DA+MS), either on the basis of the near- and mid-infrared excess detected by 2MASS and {\em WISE} or from spectroscopic evidence (Fig.\,\ref{f:dams_spectra}).   Three of these systems were visually resolved at the NTT, thus we took spectra at non parallactic angles to include both stars in the slit. One star, J1022-3226, also displays a weak H$\alpha$ emission line  that could be produced from a chromospherically active companion.  Finally, we found four DBA white dwarfs, six objects that do not present any clearly detectable spectral features (DC) at the quality of our data (with the exception of J1010-2327, which has broad features in the blue part of the spectrum that may be either due to the presence of carbon in the atmosphere or to a strong magnetic field); their spectra are shown in Fig.\,\ref{f:dbc}. 

We will not discuss the spectral classification of the hot subdwarfs and BHB stars in detail here. We roughly estimated their $T_{\rm eff}$ via a photometric fit to the NUV$BVg'r'$ photometry, using DA models with $\log{g} = 6$. We note that some of the stars have composite spectral energy distributions, which indicate the presence of a late-type companion. We display the spectra of these 25 stars in Fig.\,\ref{f:sd}, noting they span a range of temperatures between 11\,000--24\,000\,K. For the previously known stars, we advise
the reader to refer to the existing literature.

\begin{figure*}
\centering
\includegraphics[width=0.85\linewidth]{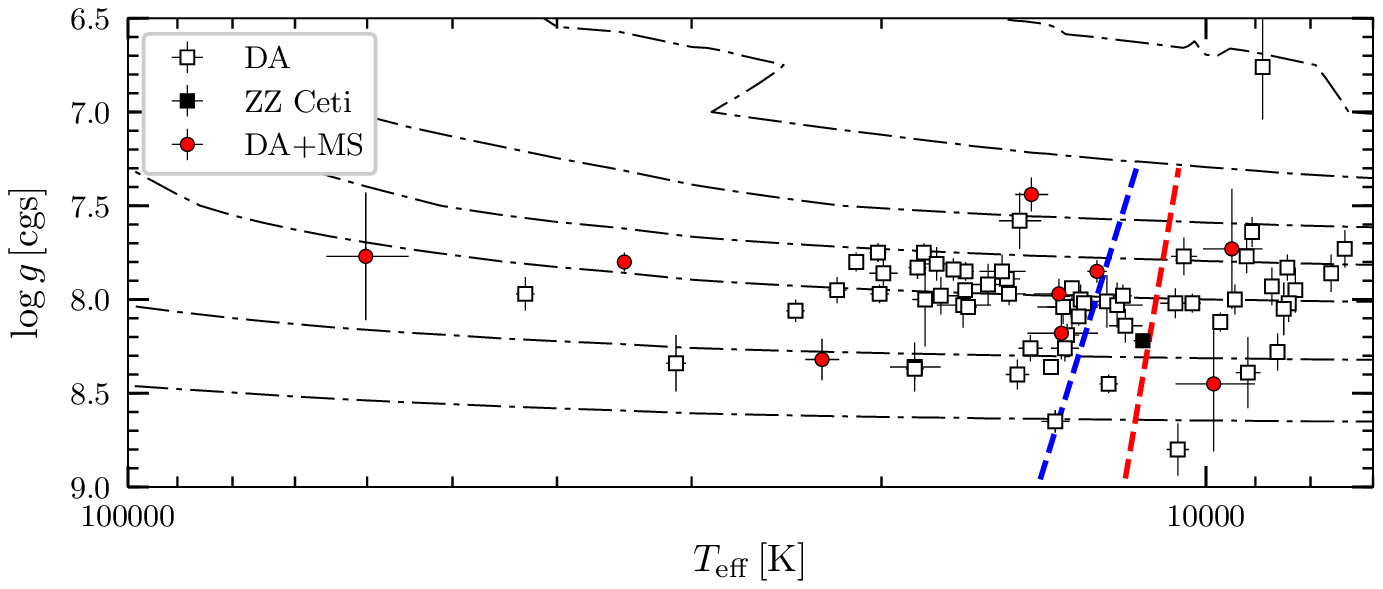}
\caption{Atmospheric parameters of DA white dwarfs. 
The \citet{Fontaine01} cooling tracks are overplotted as dash-dotted curves for 0.2, 0.3, 0.4, 0.5, 0.6, 0.8, and 1.0\,M$_{\sun}$ from top to bottom, respectively. The empirical red and blue $T_{\rm eff}$ limits of the instability strip are shown \citep[dashed lines;][]{Gianninas15}. We mark in black one confirmed ZZ Ceti star, J0502+5401.}
\label{f:atmos_da}
\end{figure*}
\subsection{Atmospheric parameters of the spectroscopically confirmed white dwarfs}

For the determination of temperature and surface gravity ($T_{\rm eff}$, $\log{g}$) of the white dwarfs we used the software {\sc fitsb2}, which estimates a best-fit via $\chi^{2}$ minimisation using a downhill simplex algorithm \citep[][]{Napiwotzki04}. The $T_{\rm eff}$ and $\log{g}$ uncertainties were estimated via a bootstrap method. We used the \citet{Koester10} model atmospheres, which adopt a ${\rm ML}2/\alpha = 0.8$ mixing-length prescription for convective atmospheres and the Stark broadening  profiles of hydrogen lines computed by \citet{Tremblay09}.

For the 60 DA single white dwarfs we fitted six Balmer lines (H$\alpha$--H8), while we excluded the H$\alpha$ and H$\beta$ from the fit for the nine DA+MS pairs when the spectral profiles were altered by the presence of the companion star.  A few examples of the best-fitting results are shown in Fig.\,\ref{f:da_lines}, illustrating the quality of both the data and the fits. We also display in Fig.\,\ref{f:da_logg} an example of how the signal-to-noise ratio in the proximity of Balmer lines and spectral resolution are the relevant factors in assessing the surface gravity of selected objects, chosen at $T_{\rm eff} \approx 10\,000$ and $T_{\rm eff} \approx 20\,000$. To estimate $T_{\rm eff}$ of non-DA white dwarfs, we used a photometric technique fitting the available ultraviolet/optical data and keeping $\log{g} = 8$ fixed.  

From the Copernico/AFOSC observations, we identified J0130+5321 (GD\,278) for which we had to arbitrarily fix $\log{g} = 6$ to obtain a satisfying fit of the spectrum and photometry. Its spectrum was acquired in good weather conditions, and presents only hydrogen lines at the resolution of the AFOSC setup.

The atmospheric parameters of the 60 single DA white dwarfs and nine DA+MS white dwarfs are listed in Table\,\ref{t:physical}. The eight stars that were previously known are: J0347$+$4358 J0502$+$5401, J0812$+$1737, J1455$+$5655, and J2233$+$2610 \citep[][]{Limoges13, Limoges15}, J0847-7312 \citep[][]{Gianninas10}, J1003$-$0337 \citep[GD\,110;][]{Kepler95}, and J1128$+$5919 \citep[][]{Girven11}. Our results 
for $T_{\rm eff}$ and $\log{g}$ agree within 2--3\,$\sigma$ with those reported in the literature.

Among the hydrogen-dominated atmospheres, there are two peculiar objects. One is a magnetic white dwarf (DAH), J2109+6507, for which we kept $\log{g} = 8$ fixed and derived its $T_{\rm eff}$ through a fit of the photometric data. The second object is a DAB white dwarf, J2047$-$1259. We obtained two spectra, the first taken at the NTT on 2014 June and the second at the INT on 2015 May, which do not show sensible differences in the relative position  and appearance of spectral features. We initially estimated a $T_{\rm eff} = 17\,000 \pm 740$\,K and $\log{g} = 9.0 \pm 0.1$, using only the Balmer lines. We then attempted a fit with a grid of DAB model spectra, which sample different [H/He] compositions. In Fig.\,\ref{f:dab}, we show the improved agreement with a synthetic spectrum of $T_{\rm eff} = 18\,630 \pm 360$\,K, $\log{g} = 8.37 \pm 0.06$, and ${\rm [H/He]} = -1$.  The quality of our radial velocity information does not exclude the possibility that this system could be a close binary pair.

\section{Physical parameters}
\label{chap5}
We estimated masses and ages for the DA white dwarfs, the DAB white dwarf, and the DA+MS white dwarfs. We interpolated the physical parameters from the Montr\'eal cooling tracks, using the spectroscopic estimates of $T_{\rm eff}$ and $\log{g}$. For $T_{\rm eff} < 30\,000$\,K, we used the carbon-oxygen core cooling models with thick hydrogen layers, which are described by \citet{Fontaine01}. For hotter stars we used the carbon-core cooling sequence of  \citet[][]{Wood95}. We applied the \citet{Tremblay13} 3D corrections to the measured atmospheric parameters of cool DA white dwarfs ($T_{\rm eff} < 15\,000$\,K), which account for the inaccurate treatment of convection in 1D model atmospheres.  The physical parameters and the 3D corrections are given in Table\,\ref{t:physical}. The distribution of hydrogen-dominated white dwarfs in the ($T_{\rm eff}$, $\log{g}$) plane is displayed in Fig.\,\ref{f:atmos_da}, where most stars are found to have $\log{g} = 8.00 \pm 0.25$. 

Seven stars have atmospheric parameters that are compatible with the instability strip of DA white dwarfs \citep[][]{Gianninas15,Tremblay15}. One is the already known ZZ Ceti star, J0502$+$540 \citep[][]{Limoges15, Green15}, for which we give more details in Section\,\ref{chap6.3}. The other six stars are candidate pulsators (cZZ in Table\,\ref{t:physical}).   

\begin{figure*}
\centering
\includegraphics[width=0.85\linewidth]{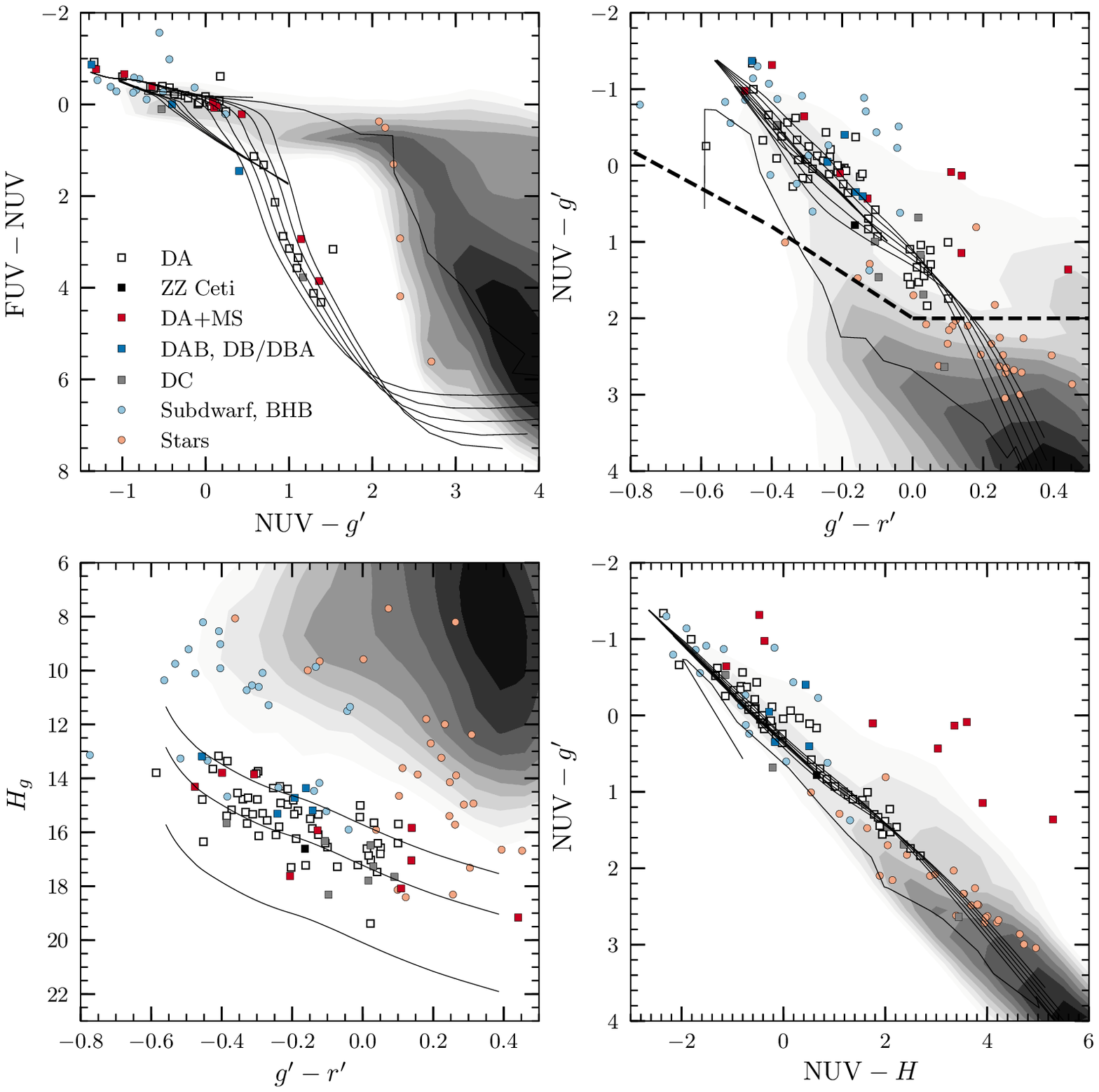}
\caption{Colour-colour and reduced proper motion diagrams of the 135 stars we have followed up. Symbols are clarified in the legend. For a description of synthetic tracks refer to Fig.\,\ref{f:samples_ccd_selection}.}
\label{f:follow-up_ccd}
\end{figure*}

Two low surface gravity stars, J0130$+$5321 (GD\,278) and J1142$-$1803, could potentially be extremely low mass white dwarfs (ELM). J0130$+$5321 would have a mass of $\approx 0.16$\,M$_{\odot}$ \citep[cf fig.\,3 in][]{Brown10}. J1142$-$1803 was previously classified as a subdwarf (sdB) by \citet{Kilkenny97}. It has narrow spectral lines as expected for a cool white dwarf ($T_{\rm eff} \approx 9000$\,K). This star has a ${\rm NUV} - g'$ colour that locates it between the white dwarf cooling tracks and the main sequence in the top left panel of Fig.\,\ref{f:follow-up_ccd}, in agreement with the low surface gravity estimate ($\log{g} = 6.76 \pm 0.28$), corresponding to $\approx 0.21$\,M$_{\odot}$.  Given that we used low-resolution gratings, we cannot rule out that at least GD\,278 is a metal-poor halo star \citep{Brown17}. Nevertheless, we note that both stars have relatively large proper motions that are compatible with those of white dwarfs.

\begin{figure*}
\centering
\includegraphics[width=\linewidth]{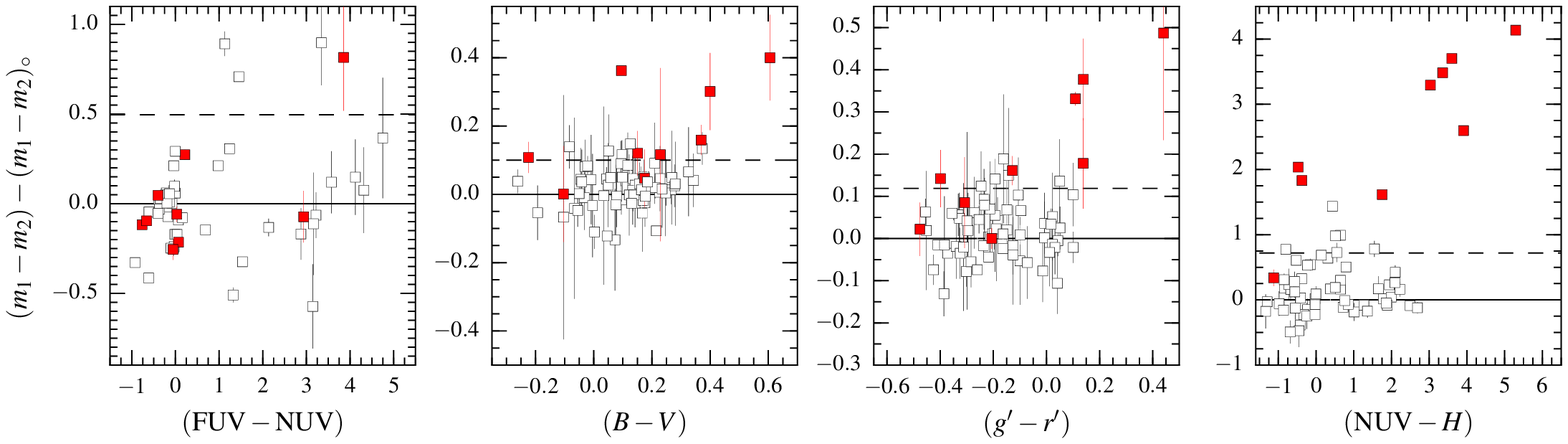}
\caption{Assessment of the colour excess for the white dwarf sample. Each panel represents a set of colours, which combine from left to right {\em GALEX}, APASS, and {\em GALEX}-2MASS bands. On the y-axis is the colour excess, i.e. the observed  minus the intrinsic colours. The dashed line marks the $E(B-V) = 0.1$ colour excess in each panel. DA white dwarfs are plotted as white squares, red symbols are used for the DA+MS pairs, which appear with strong ${\rm NUV} - H$ excesses. Generally, this spectroscopic follow-up sample has minimal interstellar reddening.}
\label{f:colours}
\end{figure*}
\section{Discussion}
\label{chap6}
\subsection{Interstellar reddening}
\label{chap6.1}
The sample of white dwarfs discussed here is made up of relatively bright, but intrinsically 
low-luminosity objects; thus, these are mostly nearby and suffer very little reddening. 
In Fig.\,\ref{f:follow-up_ccd}, we display a selection of colour-colour and reduced proper motion diagrams for the observed stars, which offer a visual comparison with the same plots for the known white dwarfs from the literature (Fig\,\ref{f:samples_ccd_selection}). The confirmed white dwarfs and subdwarfs are all located within the dashed lines that we have chosen as colour cut for the identification of white dwarf candidates in Section\,\ref{chap3.4}. The only exception is the DC white dwarf, J0201+1212 (GD\,21). The contaminants, also shown in this figure,  are mostly outside the dashed lines and share the colour space of cool ($T_{\rm eff} < 7000$\,K) white dwarfs. Finally, we stress that the presence of cool companions to white dwarfs and hot subdwarfs is clearly evident in rightmost panels of Fig.\ref{f:follow-up_ccd},  where these objects depart from their typical single-star colours.  

To confirm that the stars we observed are affected only by a small amount of interstellar reddening, we estimated the colour excess in four different colours in Fig.\,\ref{f:colours}. The leftmost panel of this figure displays the colour excess in the UV bands.  Although we attempted an empirical correction to reduce non linearity effects of {\em GALEX} photometry \citep{Camarota14},  a few hot stars (FUV$ - $NUV < 0.2\,mag) still have negative UV excess. This could be due to a wrong fit of the Balmer lines, leading us to overestimate $T_{\rm eff}$. However, in $B-V$ and $g'-r'$, most stars are consistent with $E(B-V)\le0.1$. Finally,  the NUV$-H$ colour in Fig.\,\ref{f:colours} clearly reveals the colour excess of the DA+MS systems due to the strong $H$-band contribution of their companions. We note that a few other stars follow the same trend as the DA+MS stars,  although with smaller colour excess. However, given that their 2MASS $H$ and $K_{s}$ are only upper limits, we exclude the presence of unseen companions.

\begin{figure}
\centering
\includegraphics[width=\linewidth]{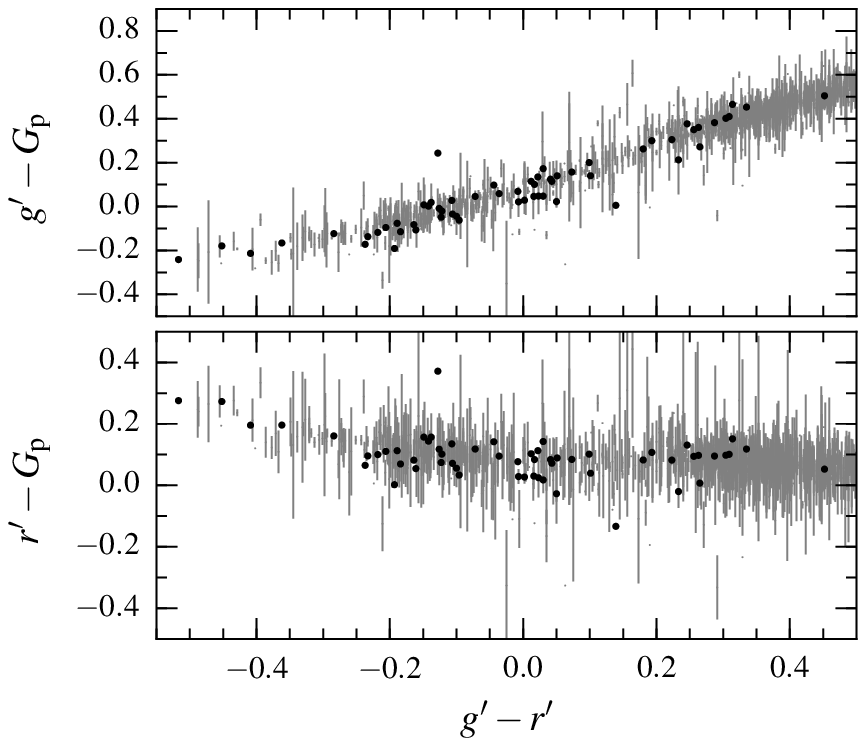}
\caption{Comparison between {\em Gaia} $G_{\rm p}$, APASS\,$g'$ and $r'$ magnitudes, plotted against $g'-r'$ colours.}
\label{f:comparison-gaia}
\end{figure}
\subsection{Comparison with {\em Gaia}}
{\em Gaia} data release 1 \citep[DR1;][]{Gaia16b} photometry became public last year. Due to known issues with the chromatic correction in the astrometry of {\em Gaia}, not all the identified sources have been released \citep[1 per cent of the total number of detected sources;][]{vanLeeuwen16}. This problem has affected especially blue objects such as white dwarfs, including the {\em Tycho/Hipparcos} objects that could have already parallaxes \citep[][]{Tremblay17}

We searched the {\em Gaia} database within 1 arcsec from the APASS coordinates of 
the 135 stars for which we have follow-up spectroscopy. We could find  $G_{\rm p}$ photometry for only 
61 of them, including 42 white dwarfs or hot subdwarfs. In Fig.\,\ref{f:comparison-gaia}, we display the comparison between {\em Gaia} and APASS magnitudes. The underlying distribution represents the photometry for the GD sample, while we overlay on top the stars we have followed up spectroscopically. About 70 per cent of the GD sample was retrieved successfully.

We find a good correlation in the range of $g' - r'$ colours covered by our sample. There are just a few outliers, and the correlations appear to depart from linearity for $g' - r' < - 0.3$; however, we identified fewer objects in this magnitude range. The most discrepant object is J1212$-$3642
(not shown in Fig.\,\ref{f:comparison-gaia}), whose APASS $g'$ and $r'$ photometry is $\approx 1$\,mag fainter with respect to {\em Gaia} $G_{\rm p}$. {\em Gaia} DR1 
contains only one source at 0.6-arcsec from the coordintates of J1212$-$3642. It does not
present detectable infrared excess or spectral peculiarities. With this information in hand, 
we suggest that APASS photometry could be systematically off.
Other two spectroscopic objects that differ by 0.2--0.3\,mag from the one-to-one correlation, J1003$-$3145 and J1438$-$3127, are DA+MS binaries for which the presence of a companion can be 
the source of photometric variability. 

\subsection{White dwarf distances}
\label{chap6.2}
We estimated spectroscopic distances by considering no extinction, but we assumed an average scatter of $E(B-V) = 0.03$ that corresponds to 0.1\,mag of extinction in the $g'$-band. We estimated the intrinsic $g'$-band magnitudes by convolving the filter profile to appropriately scaled \citep{Koester10} synthetic spectra.  We propagated the $T_{\rm eff}$ and $\log{g}$ uncertainties in order to assess the scatter in synthetic magnitudes. We found that about 12 DA white dwarfs have spectroscopic parallaxes placing them within 40\,pc from the Sun. Only J0812$+$1737 was previously known to be at less than 40\,pc from the Sun \citep{Limoges13}. Of the other 11 stars, six are in the northern hemisphere and five in the south. The nearest white dwarf we identify is J1407$-$0626, for which we estimate a distance of 25\,pc. This is a cool white dwarf ($T_{\rm eff} = 8400$\,K), which has likely escaped previous searches due to its modest proper motion ($\approx 0.1$\,arcsec/yr). Next year, the ESA {\em Gaia} mission will deliver accurate parallaxes for this object, confirming or refuting its membership to the local sample within 25~pc from the Sun.
 
The remaining DA white dwarfs are all within 100\,pc from the Sun. The most distant white dwarf in the spectroscopic follow-up sample is in the hot DA+MS system, J1022$-$3226, at 288\,pc. 

\begin{figure}
\centering
\includegraphics[width=\linewidth]{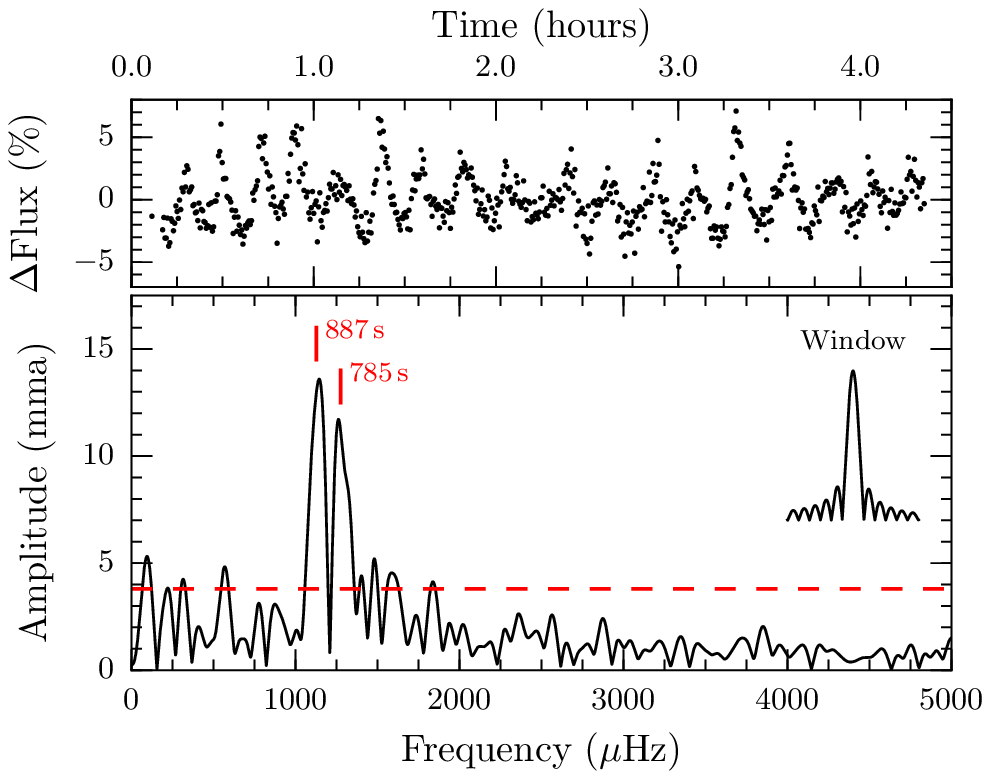}
\caption{Confirmation of pulsations for J0502$+$540. The 4.5\,hr light-curve taken at the Warwick 1-m Telescope is shown in the top panel; the discrete Fourier transform is displayed in the bottom panel. The two main peaks are labelled, and the 3-$\sigma$ threshold is displayed as a red line. In the right corner is the window function that illustrates the frequency resolution of the discrete Fourier transform.}
\label{f:WDJ0520}
\end{figure}
\subsection{ZZ Ceti stars}
\label{chap6.3}
Fig.\,\ref{f:atmos_da} identifies seven white dwarfs that fall within the instability strip of hydrogen atmospheres. One of the stars, J0502$+$540, was previously classified as ZZ\,Ceti by \citet{Limoges15} and its pulsations confirmed by \citet{Green15}. 
We re-observed this star on 2016 Jan 6 at the Warwick 1-m Telescope in La Palma for $\approx 4.5$\,hr with a 20\,s cadence. In Fig.\,\ref{f:WDJ0520} we display the Fourier transform of the light curve, where we identify two main periods in agreement with \citet{Green15}. 

Among our sample, J2233$+$2610 was also observed by \citet{Green15} and found not to be variable to above 0.07 per cent. This star is about 1000\,K hotter than the empirical instability strip; thus, it is not expected to vary. The remaining candidates will be excellent targets for {\em TESS}, which will acquire detailed light curves enabling an unprecedented level of characterisation for these and other ZZ Ceti stars. Furthermore,  we stress that the brightest stars in our sample ($g<16$) are ideal targets for ground-based observations, e.g. with Evryscope \citep{Law15} or NGTS \citep[][]{West16} that currently provide detailed time-domain photometry with a large coverage of the Southern hemisphere.

\begin{figure}
\centering
\includegraphics[width=\linewidth]{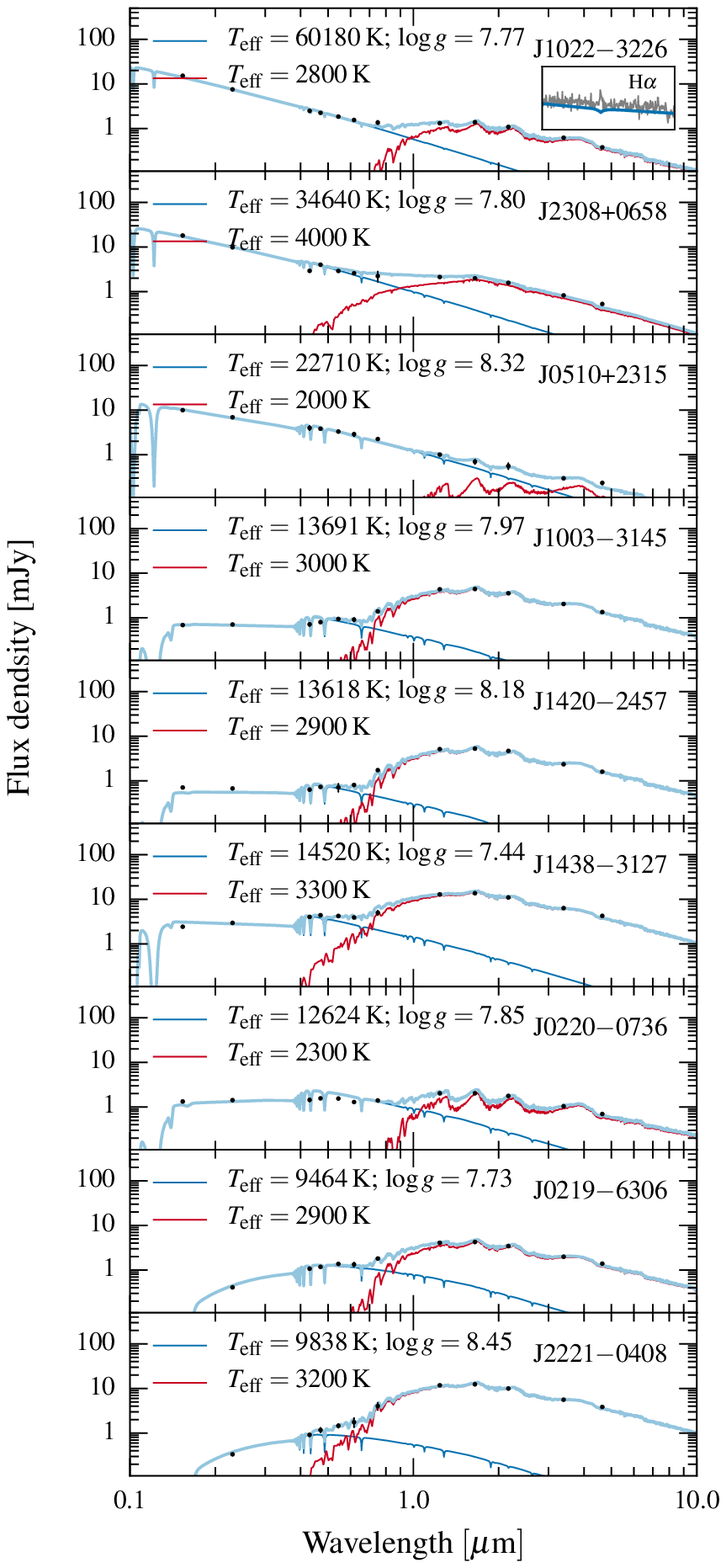}
\caption{Photometric fits of the DA+MS binaries, sorted by descending white dwarf temperature (from top to bottom). We display the white dwarf spectra (in blue), the low-mass companions spectra (in red), and the combined spectra (in cyan). We display all the photometric bands available as black dots. In the top panel, we also plot a cut-out of the H$\alpha$ region for J\,1022$-$3226, for which we detect a weak emission.}
\label{f:dams}
\end{figure}
\subsection{Late type companions}
\label{chap6.4}
We identified nine white dwarfs in our sample that display redder colours than isolated objects (Fig.\,\ref{f:colours}), which suggest the presence of late-type companion stars. We confirmed six of them based on spectroscopic evidence (e.g. composite spectra or the weak H$\alpha$  emission seen in the spectrum of J1022$-$3226; Fig.\,\ref{f:dams_spectra}).  For three of these stars, labelled in Table\,\ref{t:companions}, we could  visually resolve the companions when we acquired their spectra. Three further systems are proposed to have companions that are too cool with respect to the white dwarf to reveal their presence in optical spectra (J0520$+$2315, J0220$-$0736, and J2308$+$0658), but displaying infrared excess already in
the 2MASS $J$ band.  

We derived the $T_{\rm eff}$ of companion stars via fits to the composite spectral energy distributions (SED), including {\em GALEX}, APASS, 2MASS, and {\em WISE}~$W1/W2$ photometry. We kept the atmospheric parameters of the white dwarfs fixed to those determined from the spectroscopic analysis (Table\,\ref{t:physical}) and we built a grid of spectra for low mass stars and sub-stellar objects, taken from the latest version of the BT-Settl models \citep[][]{Allard12}. We fixed the models' metallicity to solar and $\log{g} = 5$. We derived the  intrinsic magnitudes of white dwarfs and main sequence stars by convolving  the filter profiles with the appropriate models from the \citet{Koester10} library of synthetic spectra and the BT-Settl grid, respectively. Finally, we estimated the best fits through a $\chi^{2}$ minimisation of the available photometry with three free parameters, i.e. an absolute scaling factor, the relative flux of white dwarf and companion, and the atmospheric temperature of the secondary. The results of our fits are displayed in Fig.\,\ref{f:dams}. 

We also estimated the companion masses, under the assumption they are dwarf stars. Thus, we first considered a semi-empirical initial-to-final mass relation \citep[][]{Catalan08}, from which we inferred the white dwarf progenitor mass. We then estimated the total age of the system, by approximating the stellar lifetime with a simple analytical expression \citep[e.g. ${\rm Age} = 10 \times M_{\rm prog}^{-2.5}$\, Gyr;][]{Wood92}. Finally, we determined the companion mass via interpolation of $T_{\rm eff}$ and age on the \citet[][]{Baraffe98, Baraffe03} tracks for low-mass and substellar mass objects. In Table\,\ref{t:companions}, we list the relevant physical properties of the companions ($T_{\rm eff}$, age, mass). 

We found that most of the proposed companions have masses in the range of 0.1\,M$_{\sun}$, i.e. corresponding to late-M spectral types. Spectroscopic follow-up aiming at a better characterisation of these companions is necessary, especially for the least massive of them, J0510$+$2315\,B, that we only identify through its infrared excess. For this object, we infer a mass of 0.05\,M$_{\sun}$, compatible with a brown dwarf. Although the first brown dwarf candidate to be discovered, GD\,165\,B, is a companion to a known white dwarf \citep[][]{Becklin88}, sub-stellar white dwarf companions are rare, with an occurrence rate 
of at most a few per cent, estimated from high contrast imaging techniques 
\citep[][]{Farihi05,Hogan09}, transit surveys \citep[][]{Faedi11}, 
and searches for infrared excess \citep[][]{Girven11, Steele11}. 

Another system worth mentioning is J1438$-$3127, for which we estimated a white dwarf mass of $0.36 \pm 0.04$M$_{\sun}$. A white dwarf of such a low mass could not have be formed in isolation within a Hubble time, implying it could be the product of binary interaction. \citet{Rebassa11} showed that low-mass white dwarfs are more frequent in short period (post common envelope) binaries than in the field or in wide binary systems, bringing observational evidence in support to binary interactions during the common envelope phase. Such
low-mass white dwarfs are suggested to have gone through the common envelope phase while on the first giant branch, implying that their progenitors had an initial mass of $\lesssim 1.8$\,M$_{\sun}$ \citep[][]{Zorotovic13} and consequently a lifetime  of $\gtrsim 2.3$\,Gyr. 
Given that the evolution of such systems
is complicated by the common envelope phase, we do not derive a companion mass as done for the other
binaries.

\begin{table}
\centering
\caption{Physical parameters of the low-mass companions. \label{t:companions}}
  \begin{tabular}{@{}lD{,}{\,\pm\,}{4}D{.}{.}{2}c@{}}
  \hline
Name &  \multicolumn{1}{c}{$T_{\rm eff}$} &   \multicolumn{1}{c}{Age} & Mass \\
 &  \multicolumn{1}{c}{[K]} &    \multicolumn{1}{c}{[Gyr]} & [M$_{\sun}$] \\
  \hline
J0219$-$6306\,B$^1$ &2900 , 170 & 2.05& 0.09\\
J0220$-$0736\,B     &2300  , 620 &11.70& 0.08\\
J0510$+$2315\,B     &2000  , 380& 0.40& 0.05\\
J1003$-$3145\,B     &3000  , 200 &11.55& 0.08\\
J1022$-$3226\,B     &2800  , 1050 & 1.24& 0.08\\
J1420$-$2457\,B$^1$ & 3000 , 360 & 3.37& 0.10\\
J1438$-$3127\,B$^2$ & 3300  , 140 & >2.30&  \\
J2221$-$0408\,B$^1$ & 3200  , 870  & 1.83& 0.15\\%
J2308$+$0658\,B     & 4000  , 480 & 1.80& 0.7\\
  \hline
\multicolumn{4}{l}{\scriptsize 1) Not observed at parallactic angle. Resolved binary.}\\
\multicolumn{4}{l}{\scriptsize 2) Suggested post-common envelope binary.}  
\end{tabular}
\end{table}
\section{Conclusion}
\label{chap7}
We have produced a catalogue of bright white dwarf targets for the upcoming NASA {\em TESS} 
mission as well as long-baseline photometric surveys such as Evryscope and NGTS.
The target list was compiled by cross-matching APASS, PPMXL, and {\em GALEX}. 
The catalogue includes proper motions and multi-band photometry 
for 1864 known white dwarfs brighter than 17\,mag.
Having validated our cross-match against the Lowell Observatory 
proper motion survey \citep[][]{Giclas80} and catalogues of spectroscopically confirmed white dwarfs,
we identified 427 high confidence white dwarf candidates (including 46 GD stars). We presented follow-up spectroscopy for 82 white dwarfs and 25 hot subdwarfs/BHB stars (eight white dwarfs and ten subdwarfs in the sample were previously known). Among the confirmed white dwarfs, we have identified a few interesting objects, such as a DAB white dwarf, six bright ZZ Ceti star candidates, nine WD+MS binaries, and one white dwarf at 25\,pc from the Sun. This leaves 305 high-confidence white dwarf candidates and 33 GD stars
requiring follow-up spectroscopy to ascertain their nature.

The uninterrupted light curves that {\em TESS} will obtain for the proposed stars, 
will provide unprecedented insight on their variability (ZZ Ceti stars), binary eclipses, 
as well as important statistical information on transits from exoplanets or debris around white dwarfs. The 30-min full-frame exposures of the {\em TESS} mission will increase the chances of finding serendipitous transits, by also monitoring stars that are not included in the short-cadence list of targets.

We publish the collection of photometric data for all the previously know white dwarfs that we re-identified in our analysis, which will become particularly useful with {\em Gaia} DR2 in spring 2018. Statistical analysis of spectroscopy, photometry, and parallaxes can eventually be used to obtain a very accurate estimate of cooling ages \citep[e.g.][]{OMalley13,Si17} for this sample of known white dwarfs.

\appendix
\section{Online tables}
\begin{table}
\centering
\caption{Header of the online tables containing photometry and proper motions of white dwarfs we have presented in this work. \label{t:schema}}
  \begin{tabular}{@{}lll@{}}
  \hline
  APASS & & Name based on APASS position\\
   Name & & Simbad main name\\
  ra\_apass & deg & APASS Right Ascension\\
 dec\_apass & deg & APASS Declination\\
 B      & mag & APASS\,$B$ magnitude \\
 err\_B & mag & APASS\,$B$ magnitude error\\
 V & mag & APASS\,$V$ magnitude \\
 err\_V& mag & APASS\,$V$ magnitude error\\
 g & mag & APASS\,$g'$ magnitude \\
 err\_g & mag & APASS\,$g'$ magnitude error\\
 r & mag & APASS\,$r'$ magnitude \\
 err\_r & mag & APASS\,$r'$ magnitude error\\
 i & mag & APASS\,$'i$ magnitude \\
 err\_i & mag & APASS\,$i'$ magnitude error\\
 fuv & mag & {\em GALEX} FUV magnitude \\
 err\_fuv & mag & {\em GALEX} FUV magnitude error\\
 nuv & mag & {\em GALEX} NUV magnitude \\
 err\_nuv & mag & {\em GALEX} NUV magnitude error\\
 pmra & mas/yr & $\mu_{\alpha} \cos{\delta}$ \\
 err\_pmra & mas/yr &  $\mu_{\alpha} \cos{\delta}$ error \\
 pmdec & mas/yr &  $\mu_{\delta}$ \\ 
 err\_pmdec & mas/yr &  $\mu_{\delta}$ error \\ 
 J & mag & 2MASS\,$J$ magnitude \\
 err\_J & mag & 2MASS\,$J$ magnitude error\\
 H & mag & 2MASS\,$H$ magnitude \\
 err\_H & mag & 2MASS\,$H$ magnitude error\\
 K & mag & 2MASS\,$K_s$ magnitude \\
 err\_K & mag & 2MASS\,$K_s$ magnitude error \\
 W1 & mag & {\em WISE}\,$W1$ magnitude \\
 err\_W1 & mag & {\em WISE}\,$W1$ magnitude error\\
 W2 & mag & {\em WISE}\,$W2$ magnitude \\
 err\_W2  & mag & {\em WISE}\,$W2$ magnitude error\\
  \hline
\end{tabular}
\end{table}
\section{Physical parameters of spectroscopically confirmed white dwarfs}
\begin{table*}
\vspace{-0.5cm}
\centering
\caption{{\small Properties of the follow-up white dwarf sample.
Columns are: short name of the object, based on APASS coordinates; SIMBAD name, if known;
telescope and observing date of spectrum used to assess the atmospheric parameters;
effective temperature, surface gravity, and the corresponding 3D corrections interpolated 
from the tables in \citet{Tremblay13}; masses and cooling ages, 
interpolated from \citet{Fontaine01} cooling sequences; spectroscopic distances. \label{t:physical}}}
{\tiny
\begin{tabular}{@{}p{6em}p{13em}llp{3em}D{,}{\,\pm\,}{-1}D{,}{}{2}D{,}{\,\pm\,}{-1}D{,}{}{2}D{,}{\,\pm\,}{2}D{,}{\,\pm\,}{2}D{,}{\,\pm\,}{2}@{}}
\hline
Short Name & SIMBAD &  Telescope & Date & SpT & \multicolumn{1}{c}{$T_{\rm eff}$} & 
  \multicolumn{1}{c}{$\Delta T_{\rm 3D}$} & \multicolumn{1}{c}{$\log{g}$} 
& \multicolumn{1}{c}{$\Delta \log{g}_{\rm 3D}$} & \multicolumn{1}{c}{Mass} & \multicolumn{1}{c}{$\tau_{{\rm cool}}$} & \multicolumn{1}{c}{$d$}  \\  
 
             &    &    &     & & \multicolumn{1}{c}{[K]} &  \multicolumn{1}{c}{[K]} 
	   &  \multicolumn{1}{c}{[cgs]}    & \multicolumn{1}{c}{[cgs]} 
	   & \multicolumn{1}{c}{[${\rm M}_{\sun}$]} & \multicolumn{1}{c}{[Gyr]} & \multicolumn{1}{c}{[pc]}\\
  \hline
J0008$+$3645 & GD\,4 & Copernico & 20161227 & cZZ & 12520 , 620 & -160, & 8.04 , 0.14 & -0.03, & 0.62 , 0.05 & 0.37 , 0.07 & 75 , 3 \\
J0018$-$0956 & PHL\,778 & NTT & 20140615 & DBA & 16320 , 100 &  &  8.00  &  & &  &\\
J0033$+$1021 & 2MASS\,J00335106+1021404 & NTT & 20140712 & DC & 11640 , 300 &  &  8.00  &  & &  &\\
J0106$+$3930 & GD\,10 & Copernico & 20161107 & DA & 9830 , 100 & -130, & 8.39 , 0.05 & -0.27, & 0.68 , 0.06 & 0.82 , 0.13 & 38 , 1 \\
J0126$-$6556 &  & NTT & 20140614 & DA & 13400 , 340 & 50, & 8.18 , 0.06 & 0.01, & 0.74 , 0.06 & 0.38 , 0.06 & 56 , 2 \\
J0128$-$6915 &  & NTT & 20140614 & DA & 8720 , 90 & -30, & 8.18 , 0.10 & -0.25, & 0.56 , 0.06 & 0.82 , 0.12 & 49 , 2 \\
J0130$+$2638 &  & NTT & 20140712 & DA & 13470 , 400 & 50, & 8.25 , 0.07 & 0.01, & 0.77 , 0.06 & 0.41 , 0.07 & 52 , 2 \\
J0130$+$5321 & GD\,278 & Copernico & 20161230 & DA & 8840 , 30 &   , & 6.00 , 0.00 &    , & 0.16 , 0.00 & 3.01 , 0.09 & 138 , 5 \\
J0201$+$1212 & GD\,21 & Copernico & 20161228 & DC & 7800 , 200 &  &  8.00  &  & &  &\\
J0204$+$0726 & GD\,23 & Copernico & 20161229 & DA & 18630 , 1000 &   , & 8.36 , 0.13 &    , & 0.84 , 0.06 & 0.19 , 0.05 & 98 , 4 \\
J0219$-$6306 &  & NTT & 20140615 & DAMS & 9610 , 600 & -150, & 7.99 , 0.32 & -0.26, & 0.47 , 0.05 & 0.53 , 0.10 & 70 , 4 \\
J0220$-$0736 & 2MASS\,J02202493-0736271 & NTT & 20140615 & DAMS & 12680 , 260 & -60, & 7.86 , 0.06 & -0.01, & 0.54 , 0.05 & 0.27 , 0.04 & 83 , 3 \\
J0242$-$8206 &  & NTT & 20140615 & DA & 9200 , 240 & -60, & 8.67 , 0.19 & -0.28, & 0.85 , 0.06 & 1.61 , 0.29 & 47 , 2 \\
J0347$+$4358 & Wolf\,226 & INT & 20150304 & DA & 13230 , 190 & 90, & 7.94 , 0.04 &    , & 0.58 , 0.05 & 0.26 , 0.04 & 33 , 1 \\
J0421$+$4607 &  & INT & 20150304 & DA & 7650 , 50 &   , & 8.00 , 0.10 & -0.14, & 0.53 , 0.05 & 1.06 , 0.13 & 27 , 1 \\
J0502$+$5401 & LP\,119-10 & INT & 20150305 & ZZ & 11720 , 220 & -270, & 8.33 , 0.04 & -0.11, & 0.74 , 0.06 & 0.61 , 0.10 & 41 , 2 \\
J0510$+$2315 &  & INT & 20150303 & DAMS & 22710 , 820 &   , & 8.32 , 0.11 &    , & 0.83 , 0.06 & 0.09 , 0.02 & 65 , 3 \\
J0515$+$5021 & GD\,289 & INT & 20150305 & DA & 9520 , 90 & -120, & 8.26 , 0.08 & -0.26, & 0.60 , 0.06 & 0.74 , 0.12 & 37 , 2 \\
J0529$-$0941 &  & INT & 20150305 & DA & 14550 , 370 &   , & 8.26 , 0.07 &    , & 0.77 , 0.06 & 0.33 , 0.06 & 71 , 3 \\
J0603$+$4518 &  & INT & 20150304 & DA & 16710 , 290 &   , & 7.85 , 0.05 &    , & 0.54 , 0.05 & 0.11 , 0.02 & 69 , 3 \\
J0634$+$3848 & GD\,260 & INT & 20150304 & cZZ & 12610 , 240 & -300, & 8.50 , 0.05 & -0.05, & 0.89 , 0.06 & 0.71 , 0.12 & 49 , 2 \\
J0753$+$6122 & GD\,453 & INT & 20150304 & DC & 9330 , 230 &  &  8.00  &  & &  &\\
J0812$+$1737 & GALEX J081237.8+173701 & INT & 20150306 & DA & 16730 , 140 &   , & 7.95 , 0.03 &    , & 0.60 , 0.05 & 0.13 , 0.02 & 30 , 1 \\
J0823$+$7158 & GD\,458 & INT & 20150304 & DA & 15300 , 420 &   , & 7.89 , 0.07 &    , & 0.56 , 0.05 & 0.16 , 0.03 & 88 , 3 \\
J0838$-$2146 &  & INT & 20150305 & DA & 21990 , 440 &   , & 7.95 , 0.07 &    , & 0.61 , 0.05 & 0.04 , 0.01 & 85 , 3 \\
J0847$-$7312 & WD\,0848-730 & NTT & 20140613 & DA & 17620 , 610 &   , & 7.98 , 0.10 &    , & 0.61 , 0.05 & 0.11 , 0.02 & 73 , 3 \\
J0857$-$2245 &  & NTT & 20140613 & DA & 10730 , 290 & -240, & 7.95 , 0.10 & -0.18, & 0.49 , 0.05 & 0.42 , 0.05 & 50 , 2 \\
J0859$-$3124 &  & NTT & 20140615 & DA & 31020 , 650 &   , & 8.34 , 0.15 &    , & 0.85 , 0.06 & 0.03 , 0.01 & 133 , 6 \\
J0900$-$0909 &  & INT & 20150305 & DA & 21110 , 320 &   , & 7.80 , 0.05 &    , & 0.53 , 0.04 & 0.04 , 0.01 & 73 , 3 \\
J0905$-$2342 &  & NTT & 20140615 & DA & 18270 , 250 &   , & 7.75 , 0.05 &    , & 0.51 , 0.05 & 0.07 , 0.01 & 98 , 3 \\
J0919$+$7723 &  & INT & 20150306 & DA & 9160 , 70 & -100, & 7.91 , 0.08 & -0.27, & 0.43 , 0.05 & 0.54 , 0.06 & 41 , 1 \\
J1003$-$3145 &  & NTT & 20140614 & DAMS & 13560 , 750 & 130, & 7.95 , 0.08 & 0.02, & 0.60 , 0.05 & 0.26 , 0.06 & 116 , 5 \\
J1003$-$0337 & GD\,110 & NTT & 20140614 & DA & 8420 , 50 & -20, & 8.06 , 0.07 & -0.23, & 0.51 , 0.05 & 0.79 , 0.08 & 37 , 1 \\
J1010$-$2427 &  & INT & 20150306 & DC & 20370 , 300 &  &  8.00  &  & &  &\\
J1021$-$3448 &  & NTT & 20140713 & DA & 13240 , 720 & 80, & 8.04 , 0.12 &    , & 0.64 , 0.06 & 0.31 , 0.07 & 81 , 3 \\
J1022$-$3226 &  & NTT & 20140713 & DAMS & 60180 , 5270 &   , & 7.77 , 0.34 &    , & 0.63 , 0.04 & 0.00 , 0.00 & 288 , 14 \\
J1040$-$3123 &  & NTT & 20140615 & DA & 8270 , 100 & -10, & 8.15 , 0.12 & -0.2, & 0.57 , 0.05 & 0.95 , 0.13 & 48 , 2 \\
J1052$-$1137 &  & NTT & 20140615 & DA & 18520 , 350 &   , & 7.83 , 0.06 &    , & 0.55 , 0.05 & 0.08 , 0.01 & 145 , 5 \\
J1058$+$5132 &  & INT & 20150303 & cZZ & 12320 , 640 & -230, & 8.07 , 0.12 & -0.04, & 0.63 , 0.05 & 0.40 , 0.08 & 88 , 4 \\
J1128$+$5919 & SDSS\,J112805.30+591957.9 & INT & 20150303 & DA & 14890 , 670 &   , & 7.58 , 0.15 &    , & 0.41 , 0.04 & 0.12 , 0.02 & 140 , 5 \\
J1131$-$1225 &  & NTT & 20140613 & cZZ & 12160 , 420 & -280, & 8.20 , 0.09 & -0.06, & 0.69 , 0.06 & 0.48 , 0.09 & 51 , 2 \\
J1142$-$1803 & EC\,11396-1746 & INT & 20150513 & DA & 8860 , 140 &   , & 6.76 , 0.28 &    , & 0.21 , 0.02 & 1.16 , 0.44 & 115 , 6 \\
J1212$-$3642 &  & NTT & 20140712 & DA & 20150 , 360 &   , & 7.75 , 0.05 &    , & 0.51 , 0.04 & 0.05 , 0.01 & 90 , 3 \\
J1223$-$1852 & EC\,12206-1835 & INT & 20150305 & DC & 10660 , 300 &  &  8.00  &  & &  &\\
J1229$-$0432 & LP\,675-43 & NTT & 20140712 & DC & 9950 , 400 &  &  8.00  &  & &  &\\
J1254$-$3856 &  & NTT & 20140615 & DA & 13030 , 330 & 40, & 8.01 , 0.08 & -0.01, & 0.61 , 0.05 & 0.30 , 0.05 & 87 , 3 \\
J1331$+$6809 & GD\,487 & Copernico & 20161229 & DA & 10750 , 250 & -130, & 9.00 , 0.14 & -0.2, & 1.08 , 0.05 & 1.71 , 0.21 & 36 , 2 \\
J1342$-$1413 &  & NTT & 20140615 & DA & 13870 , 410 & -70, & 8.64 , 0.06 & 0.01, & 1.01 , 0.06 & 0.75 , 0.15 & 51 , 2 \\
J1401$-$0553 &  & NTT & 20140712 & DA & 15930 , 860 &   , & 7.92 , 0.11 &    , & 0.58 , 0.05 & 0.15 , 0.04 & 107 , 4 \\
J1402$-$0736 & SDSS\,J140233.97-073650.6 & NTT & 20140615 & DA & 8590 , 100 & -10, & 8.51 , 0.10 & -0.23, & 0.78 , 0.06 & 1.56 , 0.29 & 51 , 2 \\
J1407$-$0626 &  & INT & 20150303 & DA & 8390 , 70 & -10, & 8.24 , 0.10 & -0.22, & 0.62 , 0.06 & 1.04 , 0.17 & 25 , 1 \\
J1418$+$0929 & 2MASS\,J14183529+0929195 & INT & 20150513 & DA & 17150 , 290 &   , & 7.84 , 0.06 &    , & 0.55 , 0.05 & 0.10 , 0.02 & 104 , 4 \\
J1420$-$2457 & 2MASS\,J14200652-2457026 & NTT & 20140614 & DAMS & 13530 , 1020 & 90, & 8.17 , 0.14 & 0.01, & 0.72 , 0.06 & 0.36 , 0.09 & 104 , 4 \\
J1438$-$3127 &  & NTT & 20140613 & DAMS & 14520 , 510 &   , & 7.44 , 0.09 &    , & 0.36 , 0.03 & 0.10 , 0.02 & 75 , 3 \\
J1455$+$5655 & SBSS\,1453+571 & INT & 20150304 & DA & 15230 , 520 &   , & 7.97 , 0.06 &    , & 0.60 , 0.05 & 0.19 , 0.03 & 53 , 2 \\
J1517$+$1030 & SDSS\,J151754.62+103044.3 & NTT & 20140613 & DA & 19920 , 600 &   , & 7.86 , 0.10 &    , & 0.56 , 0.05 & 0.06 , 0.01 & 122 , 5 \\
J1528$-$2515 &  & NTT & 20140613 & DA & 14960 , 370 &   , & 8.40 , 0.08 &    , & 0.86 , 0.06 & 0.38 , 0.07 & 49 , 2 \\
J1549$-$3954 &  & NTT & 20140712 & DB & 12760 , 200 &  &  8.00  &  & &  &\\
J1614$+$1046 &  & NTT & 20140613 & DA & 16800 , 970 &   , & 8.03 , 0.12 &    , & 0.64 , 0.06 & 0.15 , 0.04 & 95 , 4 \\
J1620$-$1901 &  & NTT & 20140615 & DA & 13920 , 160 & 10, & 8.35 , 0.04 & 0.01, & 0.84 , 0.06 & 0.44 , 0.07 & 47 , 2 \\
J1638$-$2035 &  & INT & 20150513 & DA & 8480 , 110 & -10, & 8.28 , 0.14 & -0.23, & 0.63 , 0.06 & 1.05 , 0.19 & 40 , 1 \\
J1700$-$8723 &  & NTT & 20140614 & cZZ & 10900 , 350 & -230, & 8.20 , 0.08 & -0.18, & 0.62 , 0.05 & 0.54 , 0.09 & 33 , 1 \\
J1729$+$0208 &  & INT & 20150513 & DA & 15460 , 760 &   , & 7.85 , 0.09 &    , & 0.55 , 0.05 & 0.15 , 0.03 & 113 , 4 \\
J1730$+$1346 & GD\,208 & INT & 20150513 & DA & 10500 , 180 & -200, & 8.27 , 0.05 & -0.25, & 0.62 , 0.06 & 0.59 , 0.09 & 39 , 1 \\
J1944$-$3603 &  & NTT & 20140615 & DA & 12970 , 150 & 10, & 8.03 , 0.03 & -0.01, & 0.63 , 0.05 & 0.32 , 0.04 & 30 , 1 \\
J1956$+$6413 &  & INT & 20150513 & DA & 17780 , 480 &   , & 7.81 , 0.09 &    , & 0.53 , 0.05 & 0.09 , 0.02 & 81 , 4 \\
J2017$-$1712 &  & INT & 20150514 & DA & 42800 , 840 &   , & 7.97 , 0.09 &    , & 0.67 , 0.05 & 0.00 , 0.00 & 103 , 4 \\
J2023$-$1115 &  & INT & 20150515 & DA & 16620 , 310 &   , & 8.04 , 0.04 &    , & 0.64 , 0.06 & 0.15 , 0.03 & 51 , 2 \\
J2025$+$7900 &  & INT & 20150516 & DA & 24030 , 460 &   , & 8.06 , 0.06 &    , & 0.67 , 0.06 & 0.03 , 0.01 & 113 , 4 \\
J2047$-$1259 & EC\,20444-1310 & INT & 20150515 & DAB & 18630 , 360 &   , & 8.37 , 0.06 &    , & 0.85 , 0.06 & 0.20 , 0.04 & 80 , 3 \\
J2101$-$0527 & SDSS\,J210110.16-052751.2 & INT & 20150516 & DBA & 50530 , 800 &  &  8.00  &  & &  &\\
J2109$+$6507 &  & INT & 20150517 & DAH & 18220 , 500 &  &  8.00  &  & &  &\\
J2112$-$2922 & EC 21096-2934 & NTT & 20140614 & DC & 10550 , 110 &  &  8.00  &  & &  &\\
J2151$-$0502 & HE 2149-0516 & NTT & 20140614 & DBA & 12240 , 30 &  &  8.00  &  & &  &\\
J2202$+$3848 & GD\,399 & Copernico & 20161230 & DA & 9270 , 60 & -100, & 8.04 , 0.09 & -0.27, & 0.49 , 0.05 & 0.60 , 0.07 & 57 , 2 \\
J2221$-$0408 & LP\,700-3/4 & NTT & 20140614 & DAMS & 9940 , 830 & -100, & 8.74 , 0.36 & -0.29, & 0.89 , 0.06 & 1.46 , 0.45 & 44 , 3 \\
J2228$-$3105 & HE\,2149-0516 & NTT & 20140615 & cZZ & 12200 , 220 & -260, & 8.02 , 0.06 & -0.04, & 0.60 , 0.05 & 0.38 , 0.05 & 60 , 2 \\
J2233$+$2610 & PM\,J22331+2610 & INT & 20150518 & DA & 13430 , 760 & 140, & 8.02 , 0.09 & 0.02, & 0.64 , 0.06 & 0.30 , 0.07 & 67 , 3 \\
J2256$-$1319 & PB\,7361 & NTT & 20140614 & DA & 20080 , 400 &   , & 7.97 , 0.05 &    , & 0.62 , 0.05 & 0.07 , 0.01 & 98 , 4 \\
J2308$+$0658 & SDSS\,J230813.05+065836.8 & NTT & 20140712 & DAMS & 34640 , 250 &   , & 7.80 , 0.05 &    , & 0.58 , 0.04 & 0.01 , 0.00 & 147 , 6 \\
J2331$-$6656 & 2MASS\,J23314533-6656078 & NTT & 20140614 & DA & 7430 , 60 & 10, & 7.86 , 0.10 & -0.13, & 0.46 , 0.05 & 0.98 , 0.10 & 43 , 1 \\
J2341$+$2924 &  & NTT & 20140712 & DA & 13120 , 230 & 10, & 8.10 , 0.05 & -0.01, & 0.67 , 0.06 & 0.34 , 0.05 & 64 , 3 \\

\hline
\end{tabular}}

\end{table*}

\section*{Acknowledgements}

This research was made possible through the use of the AAVSO Photometric All-Sky Survey (APASS), 
funded by the Robert Martin Ayers Sciences Fund.

Some of the data presented in this paper were obtained 
from the Mikulski Archive for Space Telescopes (MAST). 
STScI is operated by the Association of Universities for Research in Astronomy, Inc., 
under NASA contract NAS5-26555. Support for MAST for non-HST data is provided by the NASA 
Office of Space Science via grant NNX09AF08G and by other grants and contracts.

This publication makes use of VOSA, developed under the Spanish Virtual Observatory 
project supported from the Spanish MICINN through grant AyA2011-2.

The Digitized Sky Surveys were produced at the Space Telescope Science Institute under U.S. Government grant NAG W-2166. 
The images of these surveys are based on photographic data obtained using the Oschin Schmidt Telescope on Palomar Mountain 
and the UK Schmidt Telescope. The plates were processed into the present compressed digital form with the permission 
of these institutions. 

This research has made use of the SIMBAD database and the VizieR catalogue access tool,
operated at CDS, Strasbourg, France.

Based on observations made with ESO Telescopes at the La Silla Observatory under programme ID 093.D-0431.
We thank Javier Alarc\'on for operating the NTT.
The INT is operated on the island of La Palma by the Isaac Newton Group in the Spanish 
Observatorio del Roque de los Muchachos of the Instituto de Astrof\'{i}sica de Canarias. 
The IDS spectroscopy was obtained as part of I15AN007 and I/2015A/04.
Based on observations collected at Copernico telescope (Asiago, Italy)
of the INAF - Osservatorio Astronomico di Padova.

The research leading to these results has received funding from the
European Research Council under the European Union's Seventh Framework
Programme (FP/2007-2013) / ERC Grant Agreement n. 320964 (WDTracer).
Support for this work was provided by NASA through Hubble Fellowship grant \#HST-HF2-51357.001-A, awarded by the Space Telescope Science Institute, which is operated by the Association of Universities for Research in Astronomy, Incorporated, under NASA contract NAS5-26555.




\bibliographystyle{mnras}





\bsp	
\label{lastpage}
\end{document}